\documentclass{article}

\usepackage[utf8]{inputenc}

\usepackage[margin=2.5cm]{geometry}

\usepackage[sort&compress,numbers]{natbib}

\usepackage{float}

\usepackage{graphicx}
\usepackage{dcolumn}
\usepackage{bm}
\usepackage{hyperref}

\usepackage[dvipsnames]{xcolor}
\usepackage[normalem]{ulem}
\usepackage{amsmath}
\usepackage{amssymb,gensymb}
\usepackage{slashed}
\usepackage{float}
\usepackage{amsfonts}
\usepackage{xcolor}

\newcommand{\dd}[2]{\frac{{\rm d} #1}{{\rm d} #2}}
\newcommand{\ddd}[2]{\frac{{\rm d}^2 #1}{{\rm d}#2 ^2 }}

\title{Gravitational Waves from Phase Transitions}

\author{Djuna Lize Croon\thanks{Durham University, United Kingdom,
\href{mailto:djuna.l.croon@durham.ac.uk}{djuna.l.croon@durham.ac.uk}}
  \and
  David James Weir\thanks{Department of Physics and Helsinki Institute of
  Physics, P.O.~Box 64, FI-00014 University of Helsinki, Finland,
  \href{mailto:david.weir@helsinki.fi}{david.weir@helsinki.fi}}}

\date\today

\begin{document}

\begin{flushright}
  HIP-2024-23/TH
\end{flushright}

\begingroup
\let\newpage\relax
\maketitle
\endgroup

\begin{abstract}
    We summarise the physics of first-order phase transitions in the early
    universe, and the possible ways in which they might come about. We then
    focus on gravitational waves, emphasising general qualitative features of
    stochastic backgrounds produced by early universe phase transitions and the
    cosmology of their present-day appearance. Finally, we conclude by
    discussing some of the ways in which a stochastic background might be
    detected.
\end{abstract}

\section{Introduction}

Our understanding of the history of our universe has improved dramatically over
the past two decades. Three generations of space probes have mapped out the
electromagnetic echoes of the Big Bang -- the cosmic microwave background
radiation -- to unprecedented precision~\cite{Staggs:2018gvf}. Experiments at
the Large Hadron Collider (LHC) finally discovered the Higgs
boson~\cite{ATLAS:2012yve,CMS:2012qbp}, cementing a keystone of the so-called
Standard Model (SM) of particle physics. In recent years, we have also been able
to directly detect gravitational waves from merging black
holes~\cite{LIGOScientific:2016aoc} and neutron
stars~\cite{LIGOScientific:2017vwq}. Each of these discoveries is another
milestone in our quest to understand where we come from. Furthermore, each
discovery represents a different facet of modern physics: early universe
cosmology, experimental particle physics, and gravitational wave astrophysics.

Indeed, the earliest seconds of our universe have been the topic of speculation
and scientific inquiry since the earliest civilisations. At present, we have a
reasonably good understanding of the different stages that must have occurred,
but important open questions remain to be answered.  The aim of this present
work is to show how the three facets mentioned above, in which so much recent
progress has been made, might help us understand the universe's very earliest
moments.

\begin{figure}
    \centering

    
    \includegraphics[width=0.6\textwidth]{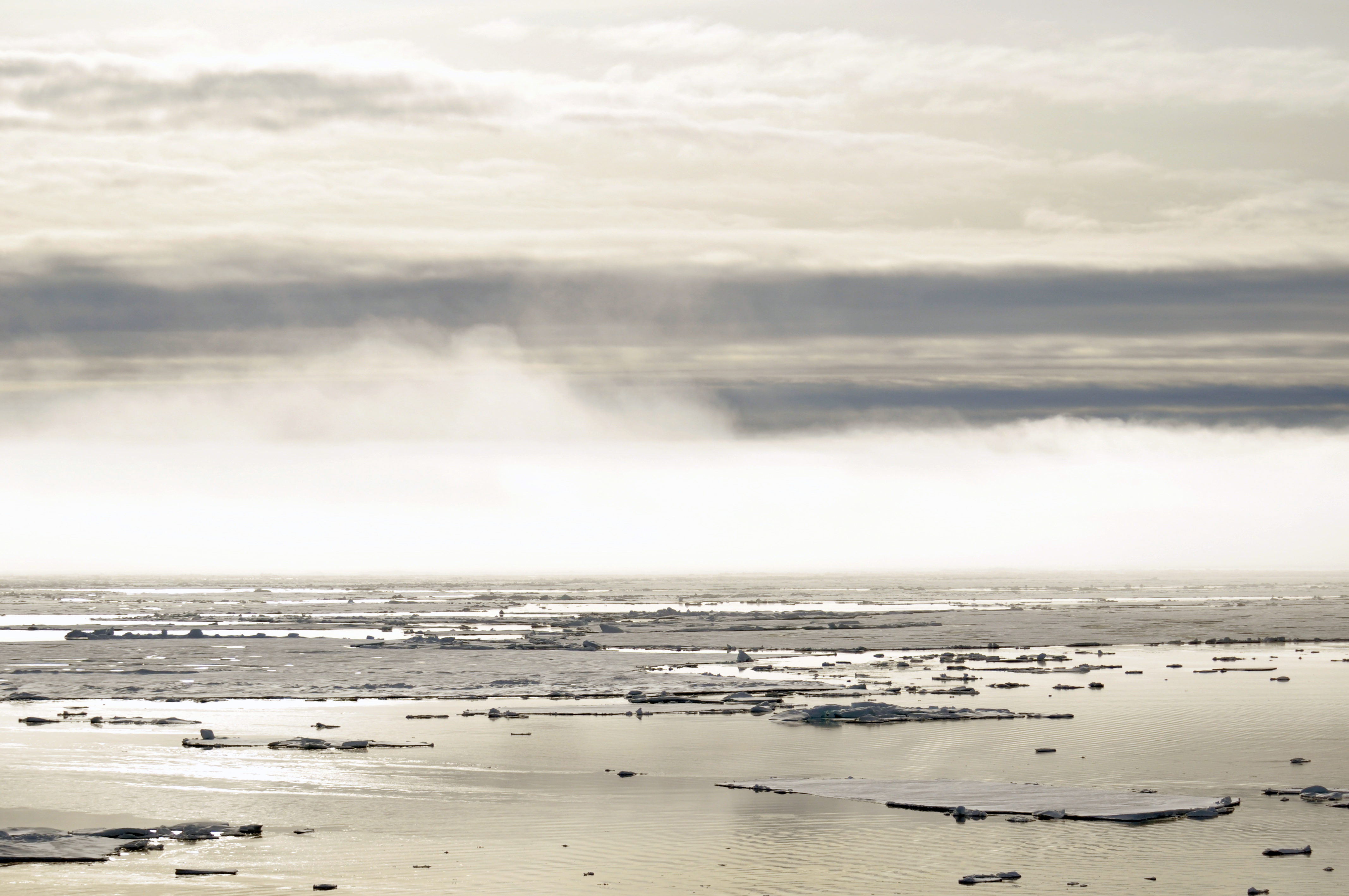}
    \caption{Fog and sea ice in the Arctic Ocean. As water vapour -- the gaseous
      phase of water -- cools, it condenses to form clouds of liquid water such
      as the fog seen here. Water, when it cools still further, freezes to form
      ice. The condensation and freezing of water are two everyday examples of
      first order phase transitions.}
    \label{fig:PhaseTransitionExample}
\end{figure}
    
As the universe expanded and cooled down, it is likely to have gone through one
or more phase transitions. In this context, a phase transition is a cosmological
processes in which the bulk properties of the very early universe changed
considerably over a relatively short time. They are analogous to the changes of
phase seen in water in everyday life: as it cools, water vapour condenses and
then freezes (see Fig.~\ref{fig:PhaseTransitionExample}) -- three phases with
very different bulk properties. In the early universe these phase transitions
leave signatures which, if detected, could help to explain some of the mysteries
that still persist, such as the imbalance between matter and antimatter in the
present universe: the observable universe is overwhelmingly made of matter, and
we are not able to explain this with currently understood physics.

Phase transitions often involve a system going out of equilibrium, and we will
see how these out-of-equilibrium phenomena could result in gravitational waves.
These gravitational waves would then travel unimpeded to our detectors, letting
us see back even further in time than the earliest detectable photons -- the
cosmic microwave background (CMB) -- which formed 380~000 years after the Big
Bang.

Collider experiments at the LHC, and other experimental particle physics
projects are looking to probe the properties of the Higgs boson -- as well as
trying to discover as-yet unseen new particles or phenomena. These might be
hitherto unseen extensions of the Standard Model of particle physics.  They
could also comprise the sought-after dark matter.

Piecing together a theory from what experimental particle physicists observe
might tell us what sort of phase transitions to expect early in the universe's
history. Conversely, detecting gravitational waves from early universe phase
transitions would inform where and how to look for new particle physics. And
both might shed some light on important open questions such as the origins of
dark matter, or the matter-antimatter asymmetry in the universe. It is this
\textsl{complementarity} between different areas of research that we want to
highlight in the current work.\footnote{See Ref.~\cite{Moss:2015fma} for a
discussion of other ways in which electroweak-scale physics can have an impact
on the universe's evolution.}

In this section we begin by reviewing the history of the universe in broad
terms, and the events which could have led to production of gravitational waves.
In Section~\ref{sec:CosmicPhaseTransitions} we will go into more detail of the
theory of early universe phase transitions, and the reasons they could have
taken place. In Section~\ref{sec:GWs} we review what gravitational waves are,
and how they can be produced by phase transitions in the early universe.  This
is followed by Section~\ref{sec:Detection}, in which we discuss ways in which
the physics of the early universe can be probed -- principally through
gravitational waves, but also colliders like the LHC. We conclude and give an
outlook in Section~\ref{sec:Conclusions}.

\subsection{Early universe physics - a brief history of what we know}
\label{sec:early}

\begin{figure}
    \centering
    \includegraphics[width=0.6\textwidth]{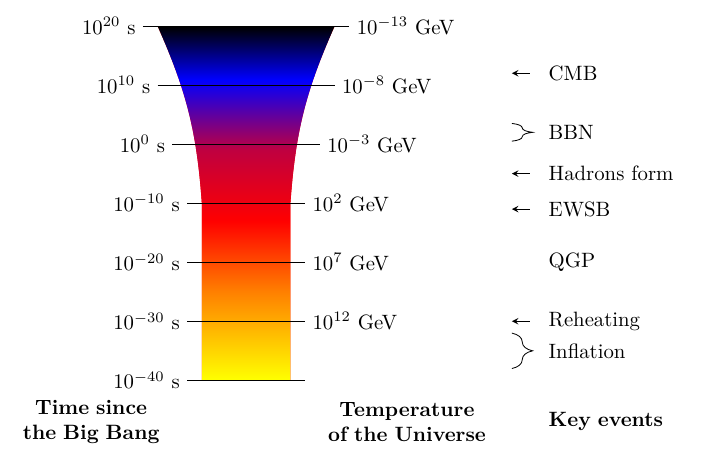}
    \caption{Chronology of the very early universe. The key events discussed in
    Section~\ref{sec:early} are shown, along a logarithmic timeline. The
    relationship between time and temperature shown here assumes that the energy
    density of radiation dominates the universe's expansion, and that there are
    no additional degrees of freedom beyond those known about in the Standard
    Model of particle physics. Various events in the history of the universe
    could have produced gravitational waves, see Table~\ref{tab:GWEvents}. }
    \label{fig:Cosmictimeline}
\end{figure}

Let us first start by reviewing what we already know about the history of the
universe, aided by Fig.~\ref{fig:Cosmictimeline}.

Our cosmic journey starts with \emph{inflation}, a period of exponential
expansion of space when the universe was between $10^{-36}-10^{-32}$ seconds old
(for an introductory review, see \cite{Liddle:1999mq}). Inflation was originally
proposed in Ref.~\cite{Guth:1980zm} to solve some of the problems with the Big
Bang model,\footnote{Namely the observation of very similar temperatures across
causally disconnected regions of the sky (the horizon problem), the extreme
flatness of the metric of space-time (the flatness problem) and the
non-observation of magnetic monopoles and other relics~\cite{Rajantie:2012xh}.}
but also explains the origin of structure in the universe: quantum fluctuations,
enhanced by inflation, can form the seeds of galaxies.

Inflation ends with a somewhat short period called \emph{reheating} in which the
universe fills up with thermal decay products of the \emph{inflaton}, a particle
which is thought to be responsible for driving the period of inflation. After
reheating, the universe contains a hot primordial soup of mostly quarks and
gluons, called the quark-gluon plasma. This lasts until the temperature of the
universe cools beyond the QCD-scale (marked ``Hadrons form" in
Fig.~\ref{fig:Cosmictimeline}), and quarks combine into \emph{hadrons}: protons
and neutrons (mostly). Through a currently undetermined process, there are more
hadrons than anti-hadrons, such that they do not annihilate immediately. This is
the big matter-antimatter asymmetry mystery mentioned above.

You will note another scale on Fig.~\ref{fig:Cosmictimeline}, above the QCD
scale, with the acronym EWSB. That stands for \emph{Electroweak Symmetry
Breaking} (an introduction can be found in Ref.~\cite{Quigg:2007fj}), and
corresponds to the scale at which the Higgs particle gave mass to the particles
of the SM. This mass mechanism is a good example of a process which can give
rise to a phase transition. We will study it in detail in
Section~\ref{sec:EWSB}.

During the era of \emph{Big Bang Nucleosynthesis} (BBN) the light elements are
produced from the protons and the neutrons (for an introduction, see
Ref.~\cite{williams1978evolution}). The plasma is still ionised at this time,
however. Besides hydrogen and helium, it contains small amounts of deuterium,
helium-3, helium-4, and lithium-7 as well. The primordial abundances of these
elements can be measured today, and are used to test theories of the early
universe.

After BBN, the primordial soup contains ionised nuclei, electrons, and photons,
until it further cools down. In the era of \emph{recombination} it becomes
energetically favoured for the electrons to bind to the nuclei to form stable
atoms. As a result, the photons are more free to travel without continuously
bumping into electrons: the universe becomes transparent. These newly released
photons are still observable today: they form the CMB, our earliest direct probe
of the universe to date

\subsection{Primordial particle physics}\label{sec:particlephysics}

As can be seen in Fig.~\ref{fig:Cosmictimeline}, the early universe was very
hot. At these temperatures, it contains a plasma of gamma rays (very energetic
photons), and other particles with a temperature dependent composition. To study
this plasma in detail, we first need to consider how particles behave at high
temperatures. The high temperature implies that the gamma rays zooming around in
the plasma can spontaneously create combinations of particles and antiparticles
with a combined rest mass smaller than their energy.  Conversely, when a
particle crosses paths with its antiparticle, they can annihilate back into a
photon. If these processes occur at the same rate, the particle species is in
thermal equilibrium with the early universe plasma.  As the temperature drops,
the average plasma energy may fall below the rest mass of a particle and its
antiparticle. When that happens, it will no longer be possible to create the
pair. We say that the particles decouple from the plasma.

The dropping temperature also affects how particles interact with other
particles, how they propagate (i.e. what their effective mass is), and sometimes
even how particles are defined. The change in behaviour of particles and
interactions as the temperature changes is very important to study the physics
of the early universe.  We know that some important things must have happened
when the universe was cooling down. Besides the processes we can confidently
mark on our cosmic timeline, some currently unknown events must have occurred,
answering open questions in particle physics.  A good example is the unknown
process producing the abundance of protons over anti-protons, which survived
until today (the proton is stable\footnote{This has been experimentally
confirmed to very high accuracy in the SM, see the baryon properties tables in
Ref.~\cite{ParticleDataGroup:2020ssz}.}): the matter-antimatter asymmetry.
Moreover, we know that some time in the first second, \emph{dark matter} must
have originated: it may have been in thermal equilibrium and then decoupled, or
it may have formed in another way. As it has eluded many dedicated experiments
today, its behaviour in the early universe may hold the key to its nature.

These two big theoretical questions -- the origins of the matter-antimatter
asymmetry, and the nature of dark matter -- motivate particle physicists and
cosmologists to study the early universe, theoretically as well as
experimentally. In the next few subsections we will explore the ways in which we
may gather information.

\subsection{Experimental information from the early universe}
\label{sec:info_early_universe}

Experimentally, we have a few different ways of learning about the early
universe. The temperature fluctuations in the photons of the CMB are the best
explored probe of observational information.  The CMB has taught us, for example, how
much dark matter there is compared to luminous matter; how many light
(relativistic) degrees of freedom there were in the early universe, and it helps
us to understand the cosmic expansion history.

The CMB comprises the earliest photons we can detect, resulting from the moment
that the universe turned from opaque to transparent (to photons).\footnote{The
corresponding distance is called the surface of last scattering.} Before this
time, the photons were in thermal equilibrium with the rest of the primordial
soup. In thermal equilibrium, temperature fluctuations get washed out, and
therefore the CMB can be seen as a snapshot of the universe at the time when
photons decoupled from the primordial plasma, 380~000 years after the Big Bang.

Nevertheless, in contrast to photons, gravitational waves were never in thermal
equilibrium with the rest of the plasma. Therefore, after a gravitational wave
is produced, it travels through space more or less unimpeded.  Therefore a
stochastic gravitational wave background (SGWB) could contain direct information
from much earlier times than the CMB does. In this way, particle physics and
cosmology are directly connected to searches for gravitational wave backgrounds.

\begin{table}[]
    \centering
    \begin{tabular}{lcccc}
        \hline \hline
        \textbf{Event} & \textbf{Time/s} & \textbf{Temp/GeV} & $g_*$ &
        \textbf{Frequency/Hz} \\
        \hline
        QCD phase transition & $10^{-3}$ & 0.1 & $\sim 10$ & $10^{-8}$ \\
        EW phase transition & $10^{-11}$ & 100 & $\sim 100$ & $10^{-5}$ \\
        ? & $10^{-25}$ & $10^9$ & $\gtrsim 100$ & $100$ \\
        End of inflation & $\gtrsim 10^{-36}$  &  $\lesssim 10^{16}$ & $\gtrsim
        100$ & $\gtrsim 10^{8}$ \\
        \hline \hline
    \end{tabular}
    \caption{Relationship between the temperature of the early universe, the
      approximate number of relativistic degrees of freedom $g_*$ (see
      Ref.~\cite{Husdal:2016haj}) at the time gravitational waves were being
      produced, and the typical frequency of gravitational waves produced
      (assuming they are produced on cosmological scales), after
      Refs.~\cite{Maggiore:2018sht,Hindmarsh:2020hop}. Assumed events in the
      Standard Model of particle physics (and the concordance model of
      cosmology) are indicated -- these models do not include events
      corresponding to the ground-based interferometer frequencies, as indicated
      by the question mark. The final column can be found by substituting the
      quantities into Eq.~(\ref{eq:generic_peak_freq}) with $B\sim 0.1$.}
    \label{tab:GWEvents}
\end{table}

If there were strong enough sources of gravitational waves in the early
universe, they could therefore be our earliest -- and highest-energy -- probes
of fundamental physics (see Table~\ref{tab:GWEvents}). Now, what types of events
could source gravitational radiation, and why do we expect them to have
happened? There are various candidate scenarios, which have a few things in
common: inhomogeneity, and the release of a lot of energy in a non-spherically
symmetric way (we will come back to this in Section \ref{sec:GWs}). Here, we
will focus on first-order phase transitions.

A phase transition is an event in which certain properties of a thermodynamic
system change. An example from particle physics is the Higgs mechanism, in which
the particles of the Standard Model (except the photon and the gluons) obtained
a mass, when our universe was about $10^{-11}$ seconds old. Phase transitions
can be classified by their \emph{order}: we are interested in first order phase
transitions, in which there are discontinuities in the properties of the system,
such as the energy density, pressure, temperature and volume. In first-order
phase transitions, latent heat is released, which will be of interest to us.

As you can imagine, there are many candidate theories in which the properties of
particles imply a phase transition happened in the early Universe. Such theories
are often related to fundamental open questions in particle physics. For
example, the question why there is more matter than antimatter in our universe
can only be answered by an early universe event which includes a departure from
equilibrium. Another open question relates to the elusive substance we call dark
matter: gravitational waves may be one of the only ways to shed light on it. We
will discuss both of these examples (and a few others) in Section
\ref{sec:models}.

In the rest of this paper we will explain the thermodynamics of a cosmic phase
transition, and give examples of classes of models in which such events occur.
After a short primer on gravitational waves, we will explain how they are
generated in a phase transition. We will finish with a survey of detection
strategies, and an outlook to the next generation of experiments. 

\section{Cosmic phase transitions}
\label{sec:CosmicPhaseTransitions}
\subsection{Spontaneous symmetry breaking}
Particle physics studies the most basic building blocks of nature. A great
triumph of the last century was the discovery that the fundamental particles we
know about, and how they interact with each other, can be organised in a simple
set of rules. The success of these rules has been demonstrated by their
prediction of new fundamental particles, which were indeed subsequently found.

The organising principles of particle physics are more commonly referred to as
symmetries. That is because of the mathematical definition of a symmetry: a
feature that remains conserved or unchanged under a transformation. For example,
if your face happened to be perfectly left-right symmetric, the transformation
of flipping your face through the vertical axis in the middle would result in
exactly the same face. An example from particle physics is the conservation of
electromagnetic charge, with the corresponding electromagnetic symmetry.
Conserved quantities also come with a force, such as the repulsive force between
two like (electromagnetic) charges.

In some cases, the symmetries of particle physics depend on the energy scale. As
the universe cools, a high-energy symmetry may not be present in the low-energy
physics. Typically, the symmetry is technically still present in the theory, but
the vacuum state does not respect it: we say the symmetry is spontaneously
broken. The spontaneous breaking of symmetries is intimately related to first
order phase transitions, so we will take a moment to review this. 

\begin{figure}[h]\centering
    \includegraphics[width=.5\textwidth]{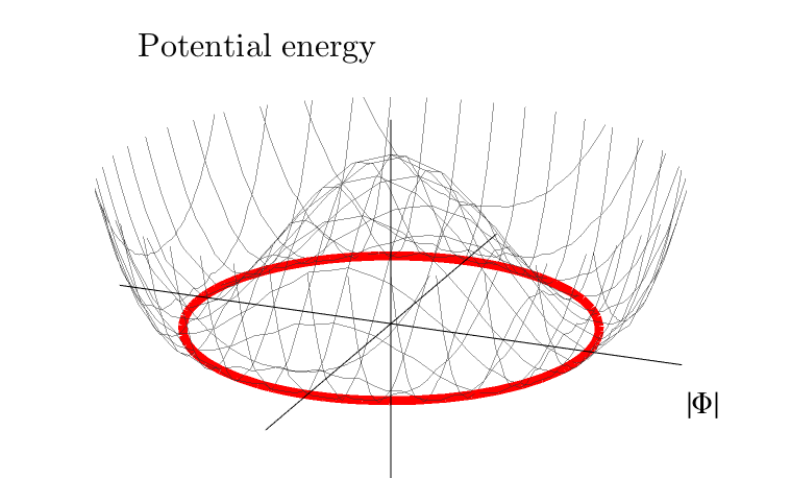}
    \includegraphics[width=.4\textwidth]{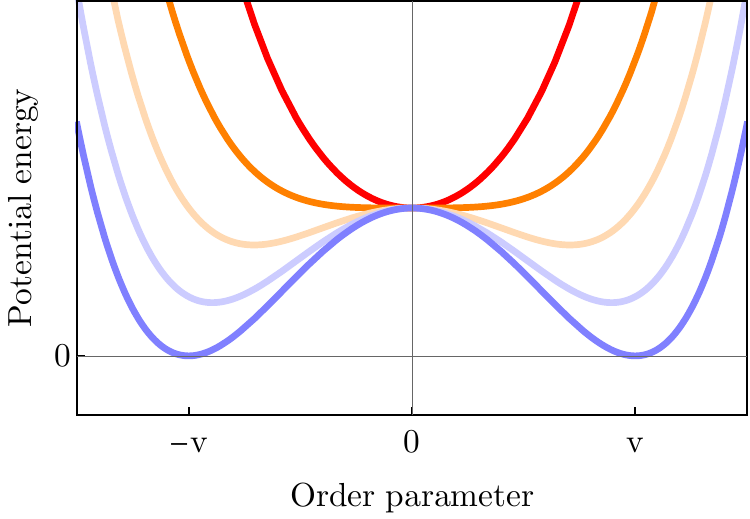}
    \caption{ Symmetry breaking potential of the Abelian Higgs model, resembling
    the underside of a wine bottle. \emph{Left:} Though the potential is
    symmetric, the vacuum (indicated in red) is not at $|\Phi| = 0$, indicating
    spontaneous symmetry breaking has taken place. \emph{Right:} Cross section
    of the potential at various temperatures: at high temperature (red line),
    the lowest energy state is symmetric about the zero axis. At low temperature
    (blue lines) the lowest energy states are not symmetric, even though the
    potential energy curve is. }
    \label{fig:abelianhiggs}
\end{figure}

As an example of a theory that features spontaneous symmetry breaking, we can
take the Abelian Higgs model in scalar electrodynamics.\footnote{In this model,
a phase transition can result in the formation of topological defects known as
cosmic strings (see e.g. \cite{Vachaspati:2015cma}), but that is not the focus
of the current work.} The potential energy of the Abelian Higgs particle is
described by,
\begin{equation}
    V(\Phi) = -m^2\Phi^\dagger \Phi + \lambda (\Phi^\dagger \Phi)^2 
\end{equation}
where $\Psi$ is complex, $\Phi = |\Phi| e^{i\alpha}$ (here $\alpha$ is a real
parameter). Clearly, $V(\Phi)$ is invariant under shifts in $\alpha$: 
\begin{equation}
    V(|\Phi| e^{i\alpha}) = V(|\Phi| e^{i(\alpha +\delta\alpha)}).
\end{equation}
You may also have noticed that the lowest value of $V(\Phi)$ (the vacuum) is not
reached at $|\Phi|=0$, but rather at $|\Phi| = v \equiv m/\sqrt{2\lambda}$, as
also demonstrated in Fig.~\ref{fig:abelianhiggs} (left). From that figure you
can also tell that the potential has rotational symmetry about the origin
(mathematically, this is called U(1) symmetry), but not about a vacuum state:
the symmetry is spontaneously broken.\footnote{You can show this mathematically
using the Abelian Higgs potential, by writing $\Phi = v + \delta \Phi$.} 

Things get more interesting when we embed our Abelian Higgs field in a plasma of
particles, and we heat that plasma up to temperature $T$. Typically, the finite
temperature implies that the Higgs particles get `kicks' from other (Higgs)
particles, which affect the potential energy in a way that at high temperatures,
to first order can be approximated as an effective mass, 
\begin{equation}\label{eq:finiteT}
    V(\Phi,T) = (-m^2+T^2)\Phi^\dagger \Phi + \lambda (\Phi^\dagger \Phi)^2.
\end{equation}
An illustration is given in the right panel of Fig.~\ref{fig:abelianhiggs} (for
ease of viewing, we show a slice in the 2d plane at, say, $\alpha=0$).  From
\eqref{eq:finiteT} we find that if $T^2\geq m^2$, the vacuum state of the theory
is indeed at $|\Phi|=0$.

Now, the picture is starting to emerge. In the hot, early universe plasma, the
temperature corrections to the potential energy of a field like the Abelian
Higgs imply that the potential has a vacuum that respects the symmetry. But as
the temperature falls below the other parameters of the potential (in this case
$m$), the vacuum shifts. This change in vacuum state is what we call a phase
transition. The Higgs vacuum value is here an order parameter: it tracks if the
phase transition has happened. 

\subsection{First order phase transitions}
In the example above, the finite temperature behaviour is clear: for $T^2 \geq
m^2$ the vacuum is at $|\Phi|=0$, for smaller temperatures, it gradually shifts
away (try it!). We call such a continuous shift in vacuum state a second order
phase transition. Second order phase transitions are relatively common in
theories of particle physics, but also relatively uninteresting: as we will see
below, to leave an imprint a sudden release of energy is needed.

\begin{figure}[h]\centering
    \includegraphics[width=.45\textwidth]{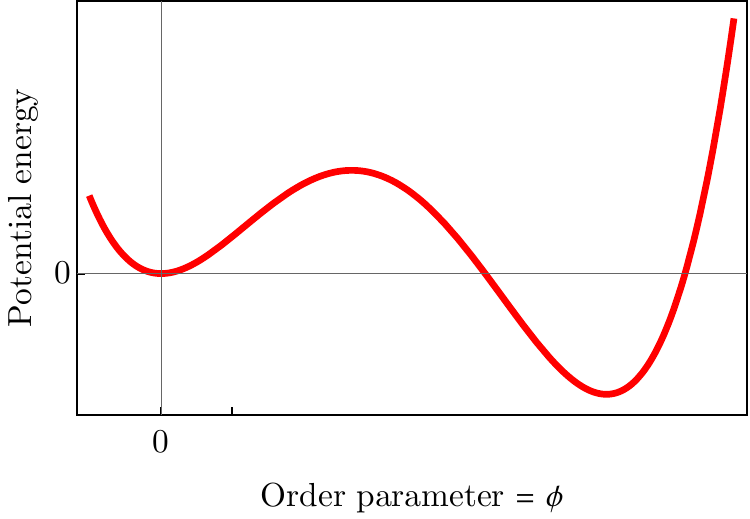}
    \caption{Form of the potential \eqref{eq:firstorderpotential} at an
    intermediate temperature $m<T<2 m \sqrt{\lambda}/\sqrt{4\lambda-1}$.}
    \label{fig:firstorderpotential}
\end{figure}

Such a sudden release of energy is a feature of a first order phase transition.
In the language of the shifting vacuum expectation value $|\Phi|$, it implies a
discontinuity. Consider for example the alternative potential energy of $\Phi$
at high temperatures,
\begin{equation}\label{eq:firstorderpotential}
    V(|\Phi|,T) = (-m^2+T^2)|\Phi|^2 -T |\Phi|^3+ \lambda |\Phi|^4,
\end{equation}
a simplified example which captures the essential behaviour of some theories we
will describe in the next section. At very high temperatures, the vacuum state
of this potential is at $\phi=0$. But for intermediate temperatures,
specifically for $m<T<2 m \sqrt{\lambda}/\sqrt{4\lambda-1}$ the lowest energy
state is away from $\phi=0$, and separated from it by a barrier. This situation
is illustrated in Fig.~\ref{fig:firstorderpotential}.

It is clear that in this type of scenario, the evolution from the symmetric to
the broken vacuum is not smooth. If it happens at all, it proceeds via the
quantum process of tunnelling through the barrier, or through a thermal
fluctuation over the barrier. In the next subsection we will outline the
formalism that captures both of these processes.

\subsection{Bubble nucleation}
\label{sec:bubble_nucleation}
To describe the phase transition, we first need to calculate the probability for
tunnelling through the potential barrier. In one-dimensional, non-relativistic
quantum mechanics, such a tunnelling probability can be found using the
semi-classical WKB approximation, which gives 
\begin{equation}
    \ddd{\psi}{x} - 2 m (V(x)-E) \psi=0,
\end{equation}
where $E$ is the energy of the state and $V(x)$ is the potential energy as a
function of position $x$ (here, as in the rest of the article, we have set
$\hbar=1$). For constant potential, this is solved by $\psi \propto \exp{(-i k
x)}$ where $k= \sqrt{2 m (E-V)}$, i.e. a plane wave, for varying potential this
can be generalised to $\psi \propto \exp{(-i \int k(x) dx)}$ with $k(x)= \sqrt{2
m (E-V(x))}$. If $\psi$ has smaller energy than the potential between points
$x_1$ and $x_2$, it is exponentially decreasing, with associated tunnelling rate
\begin{equation}
    \Gamma_{\rm WKB} \propto e^{- 
    \int_{x_0}^{x_1}
    \sqrt{2 m (V(x)-E)} dx}.
\end{equation}
It turns out that we can use something analogous in quantum field theory, the
extension of quantum mechanics to relativistic systems: a method called the
bounce method~\cite{Coleman:1977py,Callan:1977pt}. Indeed, we can go to
Euclidean space, where the theory is described by the action
\begin{equation}\label{eq:SE}
    S_E = \int d^{d}x_E \left[(\partial_{x_E} \phi)^2 + V(\phi,T) \right]
\end{equation} 
in which the subscript $E$ denotes Euclidean and the integral is over Euclidean
space. The dimension of this integral is $d=4$ for quantum tunnelling (a phase
transition happening at zero temperature) and $d=3$ for a thermal phase
transition -- in equilibrium thermal field theory in Euclidean space, the time
dimension is compact and inversely proportional to the temperature, effectively
reducing the dimensionality of problem. As usual, the equations of motion are
derived from extremising this action. Because solutions that extremise the
energy are expected to be spherically symmetric, the only relevant coordinate is
the radial one: $ r = \sqrt{x_E^2 +y_E^2 +z_E^2}$. We may therefore simplify our
equations of motion to become,\footnote{You are encouraged to derive these from
the action \eqref{eq:SE} using the Euler-Lagrange formalism.}
\begin{align} \label{eq:bounce}
    \ddd{\phi}{r} + \frac{d-1}{r}\dd{\phi}{r} &= \frac{dV(\phi,T) }{d\phi}
\end{align}
with boundary conditions,
\begin{equation}\label{eq:bounceBC}
\begin{split}
       \phi(r \rightarrow \infty)&=0, \\
    \left.\dd{\phi}{r}\right|_{r=0}&=0
\end{split}
\end{equation}
ensuring finite energy at the origin and vanishing $\phi$ at infinity. 

To find the false vacuum decay rate, the solution to \eqref{eq:bounce} with
\eqref{eq:bounceBC} can be substituted back into the Euclidean action. The decay
rate per unit volume will be depend on this result,\footnote{For further reading
on the decay rate, see
\cite{Andreassen:2016cff,Andreassen:2016cvx,Croon:2023zay}. }
\begin{equation}
    \Gamma_{} = A(T) e^{-S_E T^{4-d}}
\end{equation}
where dimensional analysis can be used to show $A(T) \sim T^4$ and $S_E $ is the
Euclidean action \eqref{eq:SE} evaluated on the solution to
Eqs.~\eqref{eq:bounce} and~\eqref{eq:bounceBC}. For thermal fluctuations,
although the solution will itself depend on temperature, it is generally the
case that the nucleation rate increases exponentially as the temperature
decreases, because the ratio $S_E/T$ decreases with temperature.

\subsection{Models with first order phase transitions}\label{sec:models}
\subsubsection{The Electroweak Phase Transition} \label{sec:EWSB} In the last
section, we discussed how phase transitions are often related to the spontaneous
breaking of symmetries. In the Standard Model of particle physics, one such
spontaneous breaking occurred in the early universe. In this phase transition,
the order parameter was the Higgs boson, and the process is responsible for
giving mass to all particles that couple to the Higgs, including itself. It is
easy to see how that works, imagine for example the Higgs boson $\phi$ coupling
to two fermionic particles (for example quarks, or electrons) $\psi$:
\begin{equation}
    \mathcal{L} \supset - y \phi \bar\psi \psi.
\end{equation}
Here $y$ is a constant and $\mathcal{L} \supset$ implies that this interaction
is part of the Lagrangian of the theory.\footnote{The bar implies a Dirac
conjugate transpose, $\bar\psi \equiv \psi^\dagger \gamma^0$ where $\gamma^0$ is
a Dirac matrix, but that is not important for the present discussion.} Now
imagine that the Higgs field changes vacuum state, from $\phi=0$ to another
value $\phi = v$. Then, we may expand $\phi = v + \delta \phi$ (where $\delta
\phi$ is a fluctuation around the vacuum), 
\begin{equation}
     \mathcal{L} \supset - ( y v \bar\psi \psi + y \delta \phi \bar\psi \psi ).
\end{equation}
The first term on the right hand side has two fermions and a constant $(yv)$, and it
defines the newly acquired mass of the fermion. The second term is again an
interaction between $\psi$ and the dynamical field $\delta \phi$.

Now that we have established that a phase transition was responsible for the
mass mechanism in the Standard Model, you may wonder what its order was. It is
possible to calculate how the theory changes with temperature, like we
demonstrated in Fig.~\ref{fig:abelianhiggs}. This will depend on all the
particles of the Standard Model that interact with the Higgs. It so happens that
with just the properties of the detected particles of the Standard Model, the
phase transition is not first order: while there is an effective $\phi^3$
interaction at finite temperatures as in \eqref{eq:firstorderpotential}, it
comes with a small coefficient, such that the barrier is not large enough to
separate the vacua at finite temperature (try putting in a small coefficient to
see that that is the case!). But the order of the phase transition could change
if new, currently undiscovered particles lurk around the corner. This is because
interactions with new particles may imply effective additional terms in the
Higgs potential, such as for example
\begin{equation}
    \mathcal{L} \supset - \frac{1}{\Lambda^2} \phi^6
\end{equation}
where $\Lambda$ has dimensions of energy, such that the Lagrangian density still
has dimensions of energy to the fourth power. With this extra effective term, a
double well potential may be realised even without a temperature-dependent
interaction. 

At this point you may object that if the existence of a gravitational wave
spectrum from the phase transition relies on the hypothesis of new particles, it
seems somewhat improbable. But there is a very good reason why cosmologists
think there was a first order phase transition in the early universe, because it
may help explain why there is more matter than antimatter in the universe, as we
briefly alluded to in Sections \ref{sec:particlephysics} and
\ref{sec:info_early_universe}. Any theory that proposes an explanation of the
matter-antimatter asymmetry must have certain features, first formulated by
Andrei Sakharov in 1967. Importantly, one of those conditions identifies the
need for out of (thermal) equilibrium dynamics. To see that that is necessary,
imagine a process that is responsible for biasing matter over antimatter - in
thermal equilibrium, this process could happen in both directions and therefore
erase the asymmetry. First order phase transitions are good examples of
reactions out-of-thermal-equilibrium, and the Higgs mechanism constitutes one of
the phase transitions we are sure did occur in the early Universe. Besides, the
Higgs interacts with the other particles of the Standard Model in such a way as
to make the creation of the matter asymmetry possible in many different models
of the phase transition. Theories of this kind are called Electroweak
Baryogenesis; electroweak after the symmetry broken by the Higgs, and
baryogenesis after the genesis of baryons (matter).\footnote{If you are
interested in reading further on this topic, we recommend
\cite{Morrissey:2012db}.}

\subsubsection{Hidden sectors}

In the last subsection, we talked about a phase transition that must have
happened in the early universe involving the particles we have discovered to
exist today, belonging to the Standard Model of particle physics. However,
answering the big open questions that still exist typically requires the
existence of additional particles, or interacting \emph{sectors} of additional
particles, which are sometimes referred to as \emph{hidden} or \emph{dark}
sectors. They are given those names because their interaction with the SM
particles (including with photons) cannot be very strong, or we would have
discovered them by now. However, we do know that a hidden sector which explains
the existence of dark matter must interact gravitationally. Therein lies the
interest in gravitational wave studies: it may be one of very few windows onto
the nature of dark matter. 

As we explained in the previous subsection, a phase transition involving the SM
Higgs boson explains the generation of mass in the early universe. This phase
transition may have been first order due to the presence of new particles. A
hidden sector has to be massive, too, to interact with gravity and explain dark
matter. So it is also possible that a corresponding dark Higgs-like mechanism
gave rise to a first order phase transition. A Higgs-like mechanism in the dark
sector is different from the SM Higgs mechanism described in the previous
subsection, because we know quite a lot about the latter: we know the Higgs mass
(at zero temperature), and we know the value of its vacuum expectation value
(again, at zero temperature). We also know how the Higgs couples to the other
particles of the SM. All these pieces of information mean that the phase
transition is quite constrained: we know it must have happened at a temperature
of about $100$ GeV, for example, implying a GW spectrum peaking at roughly
$10^{-5}$ Hz. We have also learned some things about what additional particles
could \emph{not} exist (through studies at the Large Hadron Collider, for
example), which tells us some more. None of this is true for a potential phase
transition in a hidden sector. In fact, proposed dark matter particles exist for
a vast range of masses: from $10^{-22}$~eV to macroscopic objects of the mass of
the sun, $10^{66}$~eV. There could be one such type of particle, or many, and
they could be interacting, or not. 

It may seem that with all these open questions about hidden sectors, we do not
know what we are looking for. However, as we will see in the next section,
gravitational waves from phase transitions follow a very distinctive spectrum,
independent of the microphysics that gave rise to the transition. This insight
makes it possible to study what can be learned about hidden sectors from such
transitions, which is an ongoing effort by the scientific community. 

\subsubsection{Confinement}\label{sec:confinement}

The third type of early universe phase transition we will consider is a bit
different: instead of a field like the Higgs boson obtaining a vacuum
expectation value, it occurs when particles \emph{confine}. You may have
wondered what keeps the quarks bound together inside a proton or neutron (the
collection of such things is called a hadron), and you may have heard that a
strong nuclear force is exerted between them, which increases with the distance
between the quarks. This strong force ensures that the quarks are confined
inside hadrons, inside of which they are free to move around: this is called
asymptotic freedom. 

But what does a force which increases with distance mean in the context of the
early universe? Well, in quantum field theory, short distances correspond to
higher energy. So in the very early universe, temperatures were reached which
implied that the quarks roamed around freely, in the quark gluon plasma. As the
universe cooled, the quarks started to become bound up into the hadrons, a
process we call the QCD phase transition (we mentioned it in
Table~\ref{tab:GWEvents}). It is pretty difficult to work out what the order of
that transition was, but it will probably depend on the number of free quarks
there were in the plasma: $N_f$. For the SM, $N_f=3$: three of the six quarks
(the top, the bottom, and the charm quark) are so heavy that they would have
decoupled before the phase transition.\footnote{For one (the strange quark) it
is borderline, so sometimes it is written $N_f=2+1$.} The scientific consensus
at this time, based on detailed lattice simulations~\cite{Aoki:2006we}, is that for the SM, the QCD
phase transition is not first order, but if for any reason there are more light
quarks in the plasma at the time of transition, that conclusion could
potentially be reversed. The same holds for a dark sector with a similar
confining mechanism as in QCD. So, as in the two subsections above, a first
order phase transition in the early universe is an indication that we are onto
new particle physics, beyond the Standard Model.

\section{Gravitational Waves}
\label{sec:GWs}
\subsection{Ripples in spacetime}
Gravitational waves are solutions to a wave equation, where the source term is
given by the \emph{energy-momentum tensor} of the GW-source.\footnote{See e.g.
Refs.~\cite{Flanagan:2005yc,Maggiore:2007ulw} for more detailed introductions to
this equation and the solutions discussed in this section. If you are not
familiar with index notation of linear algebra, we recommend you to review it
before reading the next sections.} The \textsl{metric tensor} $g_{\mu\nu}$ is
the key object of study in the theory of general relativity. We can expand about
some background spacetime, which we take to be flat, Minkowski spacetime with
metric $\eta_{\mu\nu}$ and consider small fluctuations $h_{\mu\nu}$:
\begin{equation}
g_{\mu\nu} = \eta_{\mu\nu} + h_{\mu\nu}, \quad \left| h_{\mu\nu} \right| \ll 1.
\end{equation}
Using Einstein's equations of general relativity, we arrive at the wave equation
for the so-called \emph{metric perturbations} $h_{\mu\nu}$:
\begin{equation}
\label{eq:metricwaveeqn}
\Box h_{\mu \nu} = - \frac{16 \pi G}{c^4} T^\text{TT}_{\mu \nu},
\end{equation}
where the $\Box$ operator is the d'Alembertian: $\Box=\partial_\mu\partial^\mu $
and where the Greek indices run over space-time dimensions: $\mu,\nu=0, 1,2,3$.
This is a second-order differential (wave) equation, with a source: $T_{\mu
\nu}$, the \emph{energy momentum tensor}. We will not show its derivation here,
but it follows directly from Einstein's field equations for general relativity.

The superscript TT denotes that we have to take the transverse-traceless part of
the energy-momentum tensor. In other words, only perturbations that are
transverse to the propagation direction and traceless in terms of the tensor
indices $\mu$ and $\nu$ source physical gravitational waves. If we consider
$h_{\mu\nu}$ to be a monochromatic travelling wave in the $x$-direction, it will
take the form 
  \begin{equation}
    h_{\mu\nu} = \left[ h_{+} \left(
      \begin{bmatrix}
        0 & 0 & 0 & 0 \\
        0 & 0 & 0 & 0 \\
        0 & 0 & 1 & 0 \\
        0 & 0 & 0 & -1
      \end{bmatrix}
      \right)
      +
      h_{\times} \left(
            \begin{bmatrix}
        0 & 0 & 0 & 0 \\
        0 & 0 & 0 & 0 \\
        0 & 0 & 0 & 1 \\
        0 & 0 & 1 & 0
            \end{bmatrix}
            \right)
            \right] e^{ikx},
  \end{equation}
  which is only nonzero in the $y$- and $z$-directions (transverse) and, as a
  matrix, has zero trace. The $h_+$ and $h_{\times}$ are the two polarisation
  components. In practice, gravitational waves from both astrophysical sources
  and the cosmological background are superpositions of polarisations,
  frequencies, amplitudes and directions of propagation -- just as is the case
  for electromagnetic radiation from astrophysical and cosmological sources.

This linear wave equation can be solved by a retarded Green's function, which
you will have come across in electrodynamics. The solution will only feature
spatial dimensions, as the temporal contribution can be related to it using
energy-momentum conservation. It is in general not very insightful - but in many
cases, we can study its behaviour by taking a limit. In particular, if the
distance $r$  to the source is large and the motion inside the source is
non-relativistic (moving at a speed much less than the speed of light, $v\ll
c$), an expansion in spherical harmonics becomes possible. The first term in
this expansion is the quadrupole, given by
\begin{eqnarray}
  \label{eq:GWquad}
h_{ab}^{\text{[quad]}} &=& \frac{2 G}{r} \ddot{Q}_{ab} \quad \quad \text{where}
  \quad \quad Q_{ab} = \int d^3 x \, \rho(t,\mathbf{x})\left( x_a x_b -
  \frac{1}{3} r^2 \delta_{ab} \right)
\end{eqnarray}
which, as we anticipated, depends only on spatial dimensions: $a,b = 1,2,3$.
$Q_{ab}$ is the \emph{quadrupole moment} of the source (the dots denote time
derivatives), and $\rho(t,\mathbf{x})$ is a time-varying mass density
distribution.

Let's pause for a moment to consider this result: gravitational waves are
generated by time-varying sources with a nonzero quadrupole moment. This is
different from electromagnetism, in which you have dipole radiation as well. An
intuitive way to see the difference is this: in electromagnetism, you have both
positive and negative charges. The dipole moment arises from these two different
charges. On the other hand, the quadrupole moment arises from the arrangement of
sources, and does not depend directly on the sign of the charges. Because mass
is the charge of the gravitational force, and mass is positive definite, there
is quadrupole radiation while dipole radiation is absent. 

The quadrupole moment describes a specific change in how the masses are
distributed around their (mutual) centre of mass.\footnote{The monopole and the
dipole describe the total amount of mass, and the distribution of mass away from
some centre in some direction. If the centre of mass is picked as the centre,
the dipole is zero.} In practise, this means:
\begin{itemize}
\item Spherically symmetric systems do not generate gravitational waves;
\item Static or uniformly moving systems do not generate gravitational waves.
\end{itemize}

\subsection{Stochastic gravitational wave backgrounds}
\label{sec:sgwbgs}

The above discussion relates to compact sources, such as what has been detected
by LIGO and Virgo to date -- and even then, it is just the leading order
contribution. For our purposes, however, we wish to think about gravitational
waves produced in the very early universe. These gravitational waves are not, in
general, produced by isolated compact objects. They would be produced by events
happening throughout the universe at around the same time. The assumptions that
go into the quadrupole formula do not hold in this context,\footnote{Influential early works like Ref.~\cite{Kosowsky:1991ua} did apply the
quadrupole approximation to colliding pairs of bubbles.} and we have to consider
the full equation \eqref{eq:metricwaveeqn}

Instead of thinking about the strain $h$ from a single source as we did in the
previous section, here the source term $T_{\mu\nu}$ comes from non-equilibrium
events happening at random anywhere within the Hubble radius (roughly speaking,
the observable universe). These will then generate a stochastic time- and
space-varying ensemble of metric perturbations $h_{ij}$. Therefore it is more
helpful to think about the \textsl{energy density} of the gravitational waves
$\rho_\text{gw}$ for a given process than the strain. Just like propagating
electromagnetic waves have energy given by the sum of the squares of the
electric and magnetic fields, the energy in gravitational waves $\rho_\text{gw}$
is given by the square of the time derivative of the metric perturbations,
\begin{equation}
  \rho_\text{gw} = \frac{1}{{32 \pi G}}\left< \dot{h}_{ij}(\mathbf{x}, t)
    \dot{h}_{ij}(\mathbf{x}, t) \right>.
\end{equation}
Here the angle brackets denote that this is an average over some spatial volume
(which must be much larger than the typical wavelength of the gravitational
waves for the energy to be well
defined~\cite{Isaacson:1968hbi,Isaacson:1968zza}). We can normalise
$\rho_\text{gw}$ to the critical energy density
\begin{equation}
  \label{eq:critdensity}
  \rho_\mathrm{c} = \frac{3H^2}{8\pi G}
\end{equation}
  in the universe at the time to get $\Omega_\text{gw}$, a cosmological density
parameter analogous to for example $\Omega_\text{rad}$ for radiation.

What is often known to cosmologists as the gravitational wave power spectrum is
the quantity $\frac{\mathrm{d}\Omega_\text{gw}(f)}{\mathrm{d} \log f}$
(sometimes shortened to $\Omega_\text{gw}(f)$) -- the scaled energy density of
the universe in gravitational waves per logarithmic frequency interval,
\begin{equation}
  \frac{\mathrm{d}\Omega_\text{gw}(f)}{\mathrm{d} \log f} = \frac{1}{\rho_\mathrm{c}} \frac{\mathrm{d}
    \rho_\text{gw}(f)}{\mathrm{d} \log f}, 
\end{equation}
where $ \frac{\mathrm{d} \rho_\text{gw}(f)}{\mathrm{d} \log f}$ is defined
through
\begin{equation}
  \rho_\text{gw} = \int_0^\infty \frac{\mathrm{d} f}{f}
  \frac{\mathrm{d} \rho_\text{gw}(f)
  }{\mathrm{d} \log f}.
\end{equation}
Note that these formulae relate to the time at which the gravitational waves
were produced, or shortly afterwards. They must be cosmologically redshifted to
today if we are to make predictions about what might be observed by
gravitational wave detectors.

In the next section we will investigate some general rules for what form
$\frac{\mathrm{d}\Omega_\text{gw}(f)}{\mathrm{d} \log f}$ can take, and how the
frequency and amplitude get redshifted to today.

\subsection{ General principles for gravitational waves from primordial physics}

In order to understand the redshifting to today, we need to understand some
cosmology. A universe which is homogeneous and isotropic (but not necessarily
static) is described by the Friedman-Lema\^{i}tre-Robertson-Walker metric, a
solution to the Einstein field equations. The dynamics of a cosmological fluid
of energy density $\rho c^2$ and pressure $p$ are described by the Friedman
equations:
\begin{align}
        \frac{\dot{a}^2 + kc^2}{a^2} &= \frac{8 \pi G \rho + \Lambda c^2}{3},   \label{eq:friedman_00} \\
        \frac{\ddot{a}}{a} &=  -\frac{4 \pi G}{3}\left(\rho+\frac{3p}{c^2}\right) + \frac{\Lambda c^2}{3}. \label{eq:friedman_trace}
\end{align}
These equations describe the evolution of $a$, the scale factor of the universe,
as a function of the energy density, pressure, curvature $k$ and the
cosmological constant $\Lambda$. The combination $\dot{a}/a \equiv H$ is also
called the Hubble factor or Hubble rate. Unless $\Lambda$ or $k$ dominate, the
Friedman equations are solved by $ a\propto t^{\frac{2}{3(1+w)}}$, where $w =
p/\rho c^2$ is the equation of state parameter of the dominant fluid (we will set $c=1$ and also assume a flat universe without a cosmological constant, meaning that
$\Lambda=k=0$ in what follows). In the early universe, the most important
cosmological fluid was relativistic and behaved like radiation, which yields $w=
1/3$ (known as the era of \emph{radiation domination}). So as time progressed,
the universe expanded. 

The expansion of the universe has implications for the particles in our plasma,
in particular for radiation. We are interested in gravitational waves produced
in the early universe. Produced at a frequency $f_*$, they get redshifted on
their way to us by the ratio of scale factors then (time $t_*$), $a(t_*)$, and
now (time $t_0$), $a(t_0)$,
giving~\cite{Maggiore:2018sht,Caprini:2018mtu,Hindmarsh:2020hop}
\begin{equation}
  f_0 = \frac{a(t_*)}{a(t_0)} f_*.
\end{equation}
We can express this relation in terms of temperature (then $T_*$, now $T_0$) by
assuming entropy, $s$, is conserved as the universe expands:
\begin{equation}
  \label{eq:scaleprop}
  s(t)a(t)^3 = \text{constant}.
\end{equation}
The first law of thermodynamics lets us relate entropy density to energy
density, pressure and temperature,
\begin{equation}
  \label{eq:entropy}
  s = \frac{\rho + p}{T},
\end{equation}
noting that $p=\rho/3$ for a relativistic fluid. Then, $\rho$ (and $p$) can be
calculated for each hot, ultrarelativistic fermion and boson degree of freedom
in the early universe plasma from the Fermi-Dirac and Bose-Einstein
distributions respectively (see e.g. Ref.~\cite{Dodelson:2003ft}), with energy
densities per degree of freedom $g$
\begin{equation}
  \label{eq:rhodofs}
  \frac{\rho_\text{bosons}}{g} = \frac{\pi^2}{30}T^4; \quad \frac{\rho_\text{fermions}}{g} = \frac{7}{8}\frac{\pi^2}{30}T^4.
\end{equation}
Defining an effective number of relativistic degrees of freedom $g_{*,s}(T)$
that accounts for the factor of $7/8$ in \eqref{eq:rhodofs} and the possibility
that some species have decoupled and are no longer in thermal equilibrium with
the others, using these expressions in \eqref{eq:entropy}
yields\footnote{Because the combination $g_{*,s}(T) T^3$ can be thought of as
shorthand for a sum over all spin and helicity states and their respective
temperatures, $g_{*,s}(T)$ is not always an integer---it will contain factors
compensating for the different prefactors of fermions and bosons, as well as the
possible different temperatures of different particle species. Furthermore, the
`s' subscript means it is specific to the entropy measurement (which carries a
$T^3$ factor).}
\begin{equation}
  s =
  \frac{2\pi^2}{45} g_{*,s}(T) T^3,
\end{equation}
and hence using \eqref{eq:scaleprop},
\begin{equation}
  \label{eq:scalefactors}
  \frac{a(t_*)}{a(t_0)} = \left( \frac{s(t_0)}{s(t_*)} \right)^{\frac{1}{3}}  =
  \left( \frac{g_{*,s}(T_0)}{g_{*,s}(T_*)} \right)^{\frac{1}{3}} \frac{T_0}{T_*}.
\end{equation}
The number of relativistic degrees of freedom has not changed since the CMB
formed, so we can set the temperature today $T_0$ as the CMB temperature today,
$2.726\, \mathrm{K}$ and the number of effective relativistic degrees of freedom
as $g_{*,s}(T_r) = 3.91$~\cite{Husdal:2016haj}. We can then express the ratio of
scale factors in terms of electronvolts, by first multiplying the temperature by
Boltzmann's constant $k_\mathrm{B}$ to get an energy, and then dividing by the
elementary charge $e$, giving $T_0 = 234.4\, \mu \mathrm{eV}$. Since we want to
express the ratio\footnote{This is purely for convenience; the equation still
works for other events in the early universe, too.} in Eq.~\ref{eq:scalefactors}
in terms of $100~\mathrm{GeV}$ (the approximate temperature when the electroweak
transition happened) and $g_{*,s} \sim 100$ (the approximate number of effective
relativistic degrees of freedom at the same time) we end up with a prefactor
\begin{equation}
  \left(\frac{3.91}{100}\right)^\frac{1}{3} \times \frac{234.4
    \times 10^{-6}}{ 100 \times 10^{9}} \approx 8.0 \times 10^{-16}
\end{equation}
giving the final equation for the redshifted frequency,
\begin{equation}
  \label{eq:freq}
  f_0  \approx 8.0 \times 10^{-16}
  \left( \frac{100}{g_{*,s}(T_*)} \right)^{\frac{1}{3}} \left( \frac{100\,
    \mathrm{GeV}}{T_*} \right) f_*.
\end{equation}
We can use this formula to work out what frequency gravitational waves produced
in the early universe will have been redshifted to today. Let us assume that
some large-scale source produces gravitational waves on some scale comparable to
the observable universe at the time in question.

Objects at a distance given by the inverse Hubble rate, $H_*^{-1}$ are receding
at the speed of light, so any causal process that produces gravitational waves
must have a characteristic wavelength $\lambda \lesssim H_*^{-1}$. We can use
\eqref{eq:friedman_00} to work out the Hubble rate. Still assuming radiation
domination, and defining another\footnote{This measure $g_*$ is specific to
measurements of energy density, carrying a $T^4$ factor compared to the $T^3$
factor of $g_{*,s}$.} effective number of degrees of freedom $g_*(T)$, the
energy density of the universe is given by
  \begin{equation}
    \label{eq:energydensityradera}
    \rho(T) = \frac{\pi^2}{30} g_*(T) T^4,
  \end{equation}
in line with \eqref{eq:rhodofs}. Remembering that we set $\Lambda=k=0$ for a
flat universe and substituting this into Eq.~(\ref{eq:friedman_00}), we find
\begin{equation}
  H_*^2 \equiv \frac{\dot{a}(t_*)^2}{a(t_*)^2} = \frac{8\pi G}{3} \rho_* = \frac{8\pi G}{3}  \frac{\pi^2}{30}
  g_*(T_*) T_*^4.
\end{equation}
Since we are working in units where $\hbar = c = 1$ and expressing energies and
temperatures in $\mathrm{eV}$, it is perhaps helpful to write
\begin{equation}
  H_* = \frac{1}{M_\mathrm{P}} \sqrt{\frac{\pi^2}{90}} \sqrt{g_*(T_*)} T_*^2,
\end{equation}
where the reduced Planck mass $M_\mathrm{P} = \sqrt{\hbar c /8 \pi G} =
2.44\times 10^{18}\, \mathrm{GeV}/c^2$.

Now we cast $H_*$ in terms of typical numbers for $g_*$ and $T_*$ at the
electroweak scale:
\begin{align}
  H_* & = \frac{1}{M_\mathrm{P}} \sqrt{\frac{\pi^2}{90}} (100)^\frac{1}{2}
  \left(\frac{ g_*(T_*)}{100}\right)^{\frac{1}{2}} (100\,
  \mathrm{GeV})^2 \left(
  \frac{T_*}{100 \, \mathrm{GeV}} \right)^2 \\
  & = 1.90\times 10^{9} \left(\frac{ g_*(T_*)}{100}\right)^{\frac{1}{2}} \left(
  \frac{T_*}{100 \, \mathrm{GeV}} \right)^2 \, \mathrm{Hz} \label{eq:finalhubble}.
\end{align}
To get to the second step, we must return to SI units. One way is to convert all
the energies and masses from electronvolts to Joules (remembering to also
multiply the Planck mass by $c^2$) and then dividing by the reduced Planck
constant $\hbar$ (with units $\mathrm{J}\,\mathrm{s}$) to get a quantity with
the right units, $\mathrm{Hz}$.

Next, suppose we take our wavelength $\lambda_*$ to be some fraction $B/H_*$ ($B
< 1$). The frequency of gravitational radiation emitted will be
\begin{equation}
  \label{eq:ourfreq}
  f_* = H_*/B.
\end{equation}
Then, we can get an expression in terms of $B$ for the frequency today by
substituting \eqref{eq:finalhubble} into \eqref{eq:ourfreq} to work out the
frequency of the gravitational waves at the time, and then using that in
\eqref{eq:freq} to redshift the resulting frequency to today. This yields
\begin{equation}
  \label{eq:generic_peak_freq}
f_0 \approx 2.7\times 10^{-6} \frac{1}{B}  \left(\frac{
  g_*(T_*)}{100}\right)^{\frac{1}{6}} \left(  \frac{T_*}{100 \,
  \mathrm{GeV}} \right) \, \mathrm{Hz}.
\end{equation}
More prosaically, this can also be thought of as `redshifting' the Hubble rate
from the time of interest to today. If a process produces gravitational waves on
length scales close to the Hubble radius, we can take $B\sim 0.1$ and obtain the
typical gravitational wave frequencies of Table~\ref{tab:GWEvents}. We will
return to the way in which the scale $B$ is set in more detail in the next
section.

Now we know the frequency at which the gravitational waves are produced, we want
to know what amplitude to expect. Gravitational waves behave like radiation, and
so their fraction of the energy in the universe scales just the same,
\begin{equation}
  \rho_{gw, 0}  = \rho_\text{gw} \left( \frac{a(t_*)}{a(t_0)}
  \right)^4.
\end{equation}

Substituting in Eq.~\eqref{eq:scalefactors} gives
\begin{align}
  \rho_{gw, 0} & = \rho_\text{gw} 
  \left( \frac{g_{*,s}(T_0)}{g_{*,s}(T_*)} \right)^{\frac{4}{3}}
\left(\frac{T_0}{T_*}\right)^4 \\
& = \rho_\text{gw} \left( \frac{g_{*,s}(T_0)}{g_{*,s}(T_*)} \right)^{\frac{4}{3}} \frac{\rho_\text{rad,0}}{\rho_\text{rad}} \frac{g_*(T_*)}{g_*(T_0)}.
\end{align}
The energy density during the era of radiation domination is given by
Eq.~\ref{eq:energydensityradera}, and we have used that in the final equality.
We divide both sides by the present day critical energy density $\rho_\text{c}
\equiv 3 H_0^2/8\pi G$, and write things in terms of the density parameters,
defined as $\Omega_\text{gw} \equiv \rho_\text{gw}/\rho_\text{c}$ for
gravitational waves, and $\Omega_\text{rad} \equiv
\rho_\text{rad}/\rho_\text{c}$ for radiation.

The critical energy density is proportional to $H_0^2$. Therefore we also
multiply by the \textsl{reduced Hubble constant} $h$, defined via the
present-day Hubble constant $H_0 = h \times 100 \, \mathrm{km} \,
\mathrm{s}^{-1} \, \mathrm{Mpc}^{-1}$, so that our results do not directly
depend on measurements of the present-day Hubble rate, and because the quoted
results for cosmological parameters $\Omega_i$ are usually of the form $h^2
\Omega_i$. We then obtain
\begin{equation}
  h^2 \Omega_\text{gw,0} = h^2 \Omega_\text{rad,0} \left( \frac{g_{*,s}(T_0)}{g_{*,s}(T_*)} \right)^{\frac{4}{3}} \frac{g_*(T_*)}{g_*(T_0)} \frac{\rho_\text{gw}}{\rho_\text{rad}}.
\end{equation}
We know the fraction of the present-day universe's energy which is
radiation,\footnote{The cosmic microwave background is very nearly a blackbody,
so the energy density parameter for photons $h^2 \Omega_\gamma$ can be
calculated given the temperature of the CMB. Then, if we know the effective
number of relativistic degrees of freedom today, $g_*(T_0)$, we can compute $h^2
\Omega_\text{rad,0} = h^2 \Omega{\gamma}/2 \times g_*(T_0)$, or apply Eq.~(1) in
Ref.~\cite{Planck:2018vyg}.} namely $h^2 \Omega_\text{rad,0} \approx 4.18 \times
10^{-5}$. Again we need $g_{*,s}(T_r) = 3.91$, but also $g_{*}(T_r) = 3.36$ (the
difference being due to the neutrinos being decoupled)~\cite{Husdal:2016haj}.

Let us assume that some fraction $C \ll 1$ of the universe's energy ends up in
gravitational waves, such that $\rho_\text{gw}/\rho_\text{rad} \approx C$ (we
are still assuming the universe is radiation-dominated, which is why we divide
by the radiation energy density). For an extremely strong first order phase
transition, we could take say $C\sim 0.1$ (which would be possible if the source
were very long-lasting and very efficient).  
Many processes, including phase transitions, produce gravitational wave spectra
that have a clear peak at some frequency, so let us also consider an idealised
power spectrum with all the power concentrated close to one frequency $f_*$,
\begin{equation}
  \frac{\mathrm{d}\Omega_\text{gw}(f)}{\mathrm{d} \log f} = \frac{\rho_\text{gw}}{\rho_c} S(f,f_*).
\end{equation}
It turns out that the simplest spectral shape $S(f,f_*)$ which has a single peak
and is consistent with physical requirements is a \textsl{broken power law} (a
Gaussian would tend to a constant for $f \ll f_*$, and a Dirac delta function
would offer less physical insight). Furthermore, a broken power law is quite
close to what simulations and modelling predict for a first-order phase
transition. We take~\cite{Caprini:2009fx}
\begin{equation}
  S(f,f_*) = \frac{\left(f/f_*\right)^3}{1+\left(f/f_*\right)^4}.
\end{equation}
As explained in Ref.~\cite{Caprini:2009fx}, this ansatz also satisfies important
physical arguments. It is causal (it looks like white noise at large
wavelengths, i.e. small $f$, because it goes as $f^3$) and means that there is
finite total energy in gravitational waves, i.e. $S(f,f_*) \to 0$ as $f\to
\infty$. We have chosen here that $S(f) \propto f^{-1}$ at large $f$ but this
depends on the details of the system (see Section~\ref{sec:fopt}).

\begin{figure}[tb]
    \centering
    \includegraphics[width=.7\textwidth]{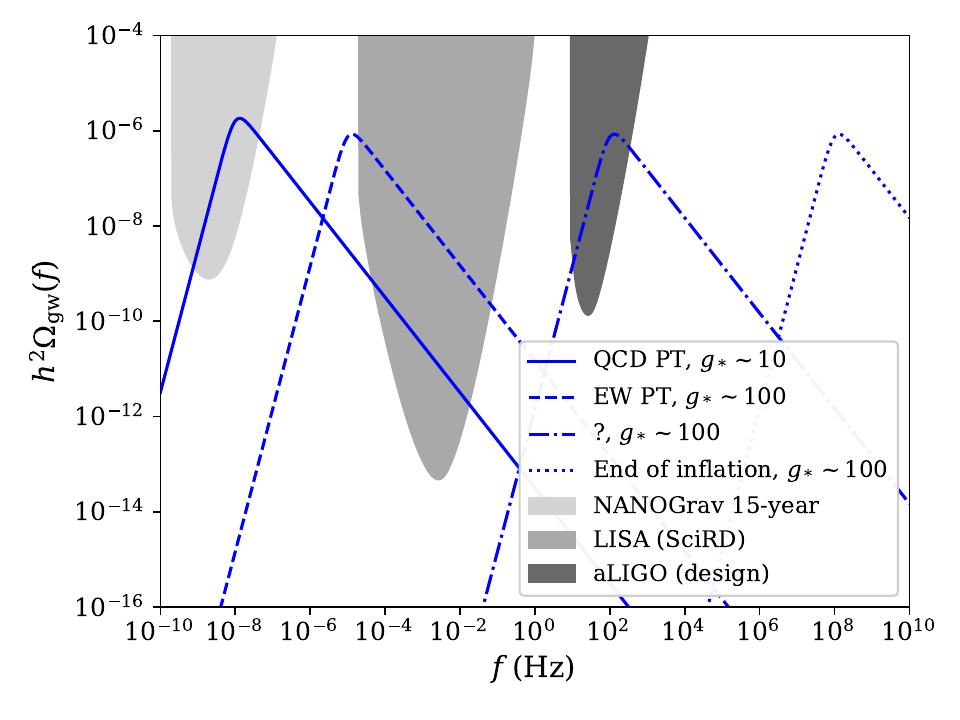}
    \caption{Best-case scenarios for gravitational wave power spectra, in the
    cases given in Table~\ref{tab:GWEvents}. Note that setting $C=0.1$ is a
    simplistic approach to computing the anticipated amplitude of the different
    scenarios, but that otherwise there is a very mild dependence on the number
    of relativistic degrees of freedom. Also shown are the power-law integrated
    sensitivity curves~\cite{Thrane:2013oya} anticipated for advanced LIGO and
    for LISA~\cite{Babak:2021mhe}, each assuming a 5 year observation time, and
    for NANOGrav~\cite{NANOGrav:2023ctt} based on the 15 year data release. If a
    background signal has a power-law shape and intersects the shaded regions,
    then in the absence of astrophysical foregrounds, it may be detectable. Our
    sensitivity curves assume a signal-to-noise ratio of 5.}
    \label{fig:bestguess}
\end{figure}

Then, taking $g_{*,s}(T_*) \approx g_*(T_*)$ and casting in terms of typical
electroweak-scale numbers (so that they are both close to 100), the amplitude of
gravitational waves today (when the frequency $f_*$ has become $f_0$) is
approximately
\begin{equation}
  h^2 \frac{\mathrm{d}\Omega_\text{GW,0}(f)}{\mathrm{d} \log f}  \approx 4.18\times 10^{-5} \times \frac{3.91^{\frac{4}{3}}}{3.36} \times  100^{-\frac{1}{3}}  \left( \frac{100}{g_{*}(T_*)} \right)^{\frac{1}{3}} C S(f,f_0)
\end{equation}
and with our broken power law spectral shape for $S(f,f_0)$ and $C\sim 0.1$,
\begin{equation}
  h^2 \frac{\mathrm{d}\Omega^\text{ideal}_\text{GW,0}(f)}{\mathrm{d} \log f}
   \approx 1.65 \times 10^{-6} \left(\frac{100}{g_*(T_*)} \right)^{\frac{1}{3}} \frac{\left(f/f_0\right)^3}{1+\left(f/f_0\right)^4}.
\end{equation}
We can use this last equation and the values in Table~\ref{tab:GWEvents} to plot
generic highest-possible amplitudes for primordial stochastic backgrounds at
different energy scales (Figure~\ref{fig:bestguess}). Note how the peak
amplitude of the redshifted present-day power spectrum is only weakly dependent
on the time in the history of the universe at which the transition took place,
\textsl{assuming} a given fraction of the universe's energy participates in the
transition.

We will see in Section~\ref{sec:Detection} that the detectability of a signal
depends on the signal-to-noise ratio (SNR), not the raw amplitude, which in turn
depends on the details of how the signal is being observed. This is because the
SNR comes from integrating the power spectrum over all frequencies, and also
integrating over the time for which the detector will be sensitive to the
source. The sensitivity curves in Figure~\ref{fig:bestguess} are so-called
`power-law integrated sensitivity curves', giving an indication of the typical
amplitude of gravitational wave background that might be detectable.

The results of this section should apply to any process in the early universe
that sourced gravitational waves on large scales. In the next section, we will
discuss the physics of gravitational wave production after a first-order phase
transition in particular.

\subsection{Gravitational waves from a first order phase transition}
\label{sec:fopt}

We have seen already that a first-order phase transition involves bubbles of the
new phase nucleating in space, expanding and merging together. We also pointed
out that isolated spherical objects don't source gravitational waves, because
they don't have any terms beyond the monopole when expanded in terms of
spherical harmonics. The bubbles, as they expand, will generally create a
reaction front in the plasma around them -- if the particles which make up the
plasma are coupled to the field undergoing the phase transition. The heated
plasma inside the reaction front carries a lot of the energy released as latent
heat during the transition, but we assume that this shell is also
spherical.\footnote{The hydrodynamical behaviour of this reaction front may be
such that \emph{instabilities} form, altering its shape away from an ideal
sphere. However, this will happen on length scales many orders of magnitude
shorter than the final bubble radius~\cite{Megevand:2013yua}.} On the other
hand, if there is no such plasma, or it is only very weakly coupled to the field
undergoing the phase transition, then the released energy goes into accelerating
the bubble walls to ultrarelativistic speeds. We will return to this in
Section~\ref{sec:vacuumrunaway}.

\begin{figure}[h]
    \includegraphics[width=\textwidth]{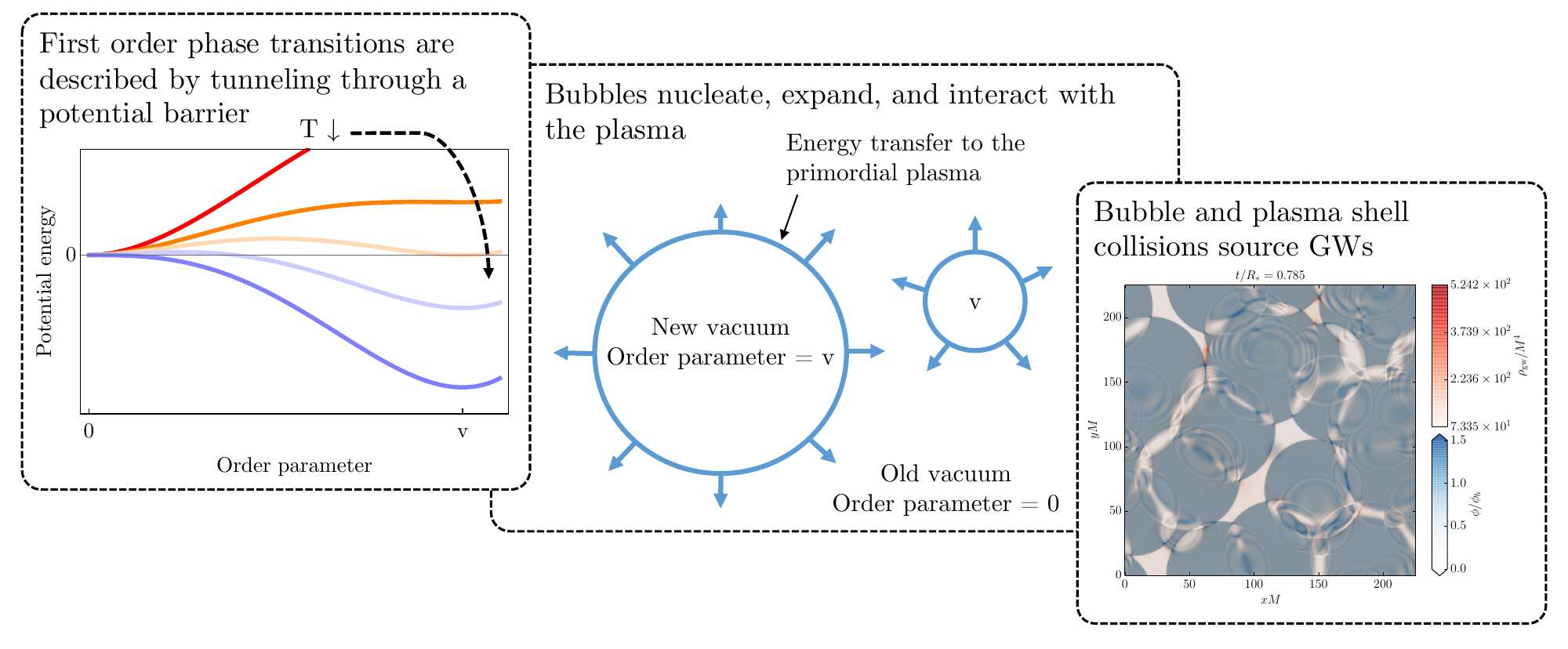}
    \caption{ Illustration of the generation of gravitational waves in a first
    order phase transition. We thank Daniel Cutting for the simulation snapshot,
    based on work in Ref.~\cite{Cutting:2018tjt}. }
    \label{fig:schematicPTgraphic}
\end{figure}

The bubbles and associated reaction fronts continue to expand, and (if the
universe is expanding slowly enough) eventually collide with one another (see
Fig.~\ref{fig:schematicPTgraphic}). At this point, the system starts to produce
gravitational waves. The heated shells of plasma may (if the transition is not
so strong) just pass through one another or (if the transition is stronger)
interact nonlinearly with one another. If the heated plasma is mostly in front
of the bubble wall (termed a deflagration), these interactions can raise the
temperature and pressure and potentially slow the bubble wall down. Eventually,
however, the bubbles will collide, leaving the whole universe in the new phase.

Both the colliding (and disappearing) scalar field bubbles and the overlapping
shells of plasma are sources of gravitational waves. Even if the reaction fronts
pass through each other without interacting strongly, they source gravitational
waves through the stress-energy tensor for the plasma, which is proportional to
fluid 4-velocity $U^\mu$ squared:
\begin{equation}
  \label{eq:fluidshearstress}
    T^{\mu\nu}_\text{(f)} = (e + p)U^\mu U^\nu,
\end{equation}
where $e$ is the energy and $p$ the pressure in the plasma, and $U^\mu$ is its
four-velocity. This then forms the source on the right hand side of
Eq.~\eqref{eq:metricwaveeqn}.

\begin{figure}
  \centering
  \includegraphics[width=0.7\textwidth]{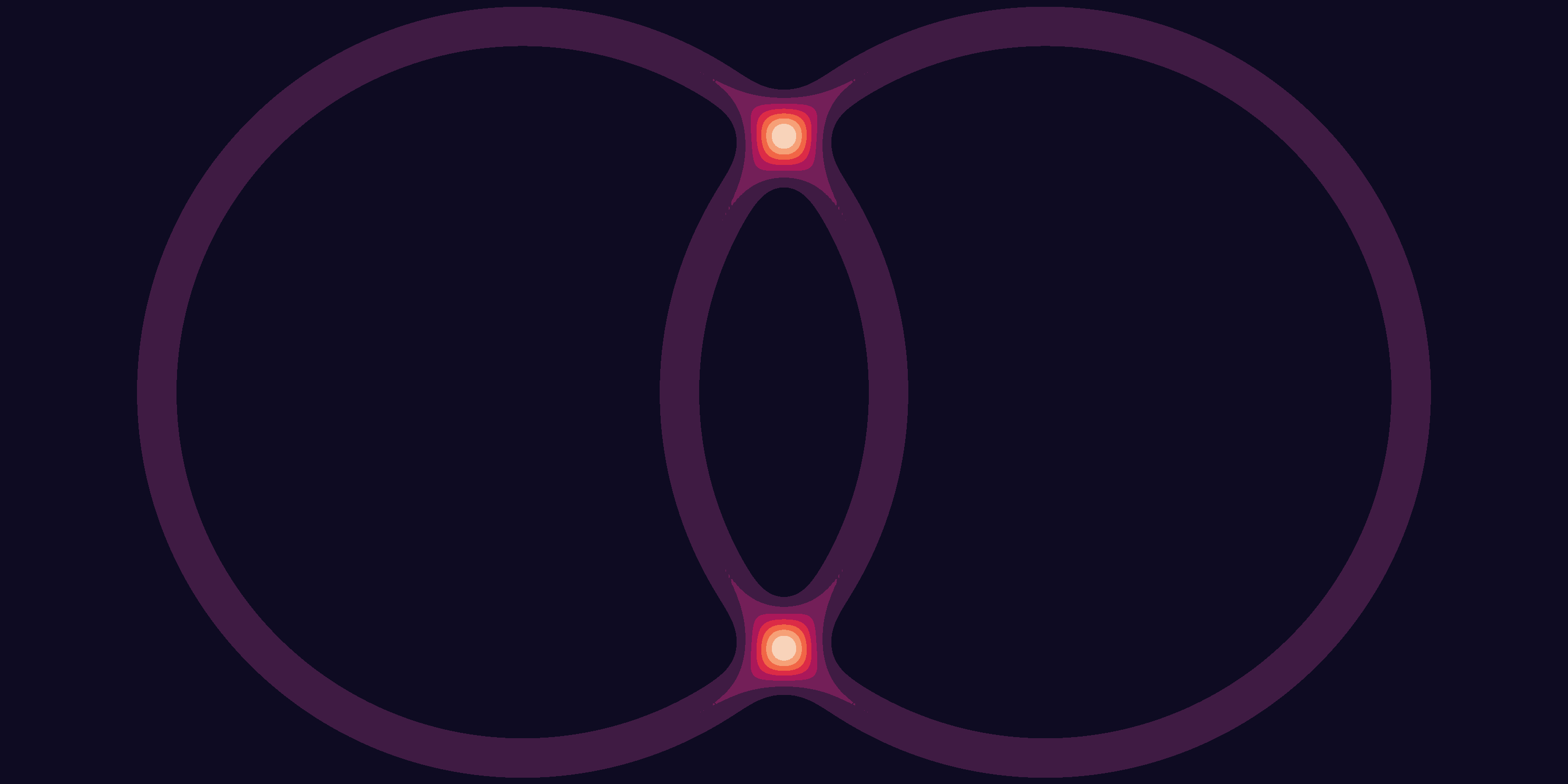}
    \caption{Illustration of the square of the 4-velocity for two overlapping
    plasma shells. }
    \label{fig:shells}
\end{figure}

The general principle that led to Eq.~\eqref{eq:GWquad} also applies to
primordial gravitational waves -- the system has to have a nonvanishing
quadrupole moment in order to source gravitational waves. In practice, this
happens when the bubbles and shells of plasma associated with them start to
overlap. Note how the fluid velocity four-vector appears twice in
Eq.~\eqref{eq:fluidshearstress}. When the plasma shells overlap, the system
cannot be decomposed into spherical shells with the same squared fluid velocity
(see Fig.~\ref{fig:shells}): the source is no longer spherically symmetric, the
quadrupole moment is no longer zero, and gravitational waves are produced.

After the bubble walls have gone, the reaction fronts are free to move, and
overlap with each other. Without a driving force provided by the latent heat of
the transition, however, they travel at the propagation speed for
perturbations in a plasma: the speed of sound. Indeed, one can at this point
think of the remnants of the reaction fronts as \emph{shells of sound waves}.
The shells can go on expanding, and overlapping with other shells, and sourcing
gravitational waves---and, if the transition is strong enough, interacting
nonlinearly with the other shells, too.

This goes on for a relatively long time, until either the universe expands
enough for the energy in the sound waves to be attenuated, or nonlinearities in
the plasma become significant. By dimensional analysis, this will happen on a
timescale proportional to the most important length scale in the system divided
by a typical velocity. Conventionally, this is taken to be the ratio of the
bubble radius at collision $R_*$ to the (enthalpy-weighted) root mean square
4-velocity $\overline{U}_\mathrm{f}$,
\begin{equation}
  \label{eq:nltimescale}
    \tau_{\text{nl}} = \frac{R_*}{\overline{U}}.
\end{equation}
This is typically called the \textsl{nonlinearity time} as the main thing which
happens is the fluid field $U^\mu$ develops nonlinearities on this scale.
Notably, these will include discontinuities in the fluid termed \emph{shocks}
(in strong transitions, shocks can also be present at the leading or trailing
edge of reaction fronts, but these will typically disappear after the
transition). Other nonlinear behaviour, such as the establishment of turbulence,
can also take place.

In most milder scenarios, however, the overlapping sound shells are the main
source of gravitational waves. Based on computer
simulations~\cite{Hindmarsh:2013xza,Hindmarsh:2015qta,Hindmarsh:2017gnf} (see
Figure~\ref{fig:schematicPTgraphic}, right, for an example of a simulation
configuration) an ansatz was developed for the gravitational wave power
spectrum,
\begin{equation}
  \label{eq:gwansatz}
  h^2 \Omega_\text{sw}(f) =
  1.19 \times 10^{-6} \left(\frac{100}{g_*(T_*)} \right)^{\frac{1}{3}}
  \Gamma^2 \overline{U}_\mathrm{f}^4 \left(\frac{H_*}{\beta}\right)
  v_\mathrm{w} \, S_\text{sw}(f, f_0).
\end{equation}
Here, $\beta = (8\pi)^{1/3} v_\mathrm{w}/R_*$, $v_\mathrm{w}$ is the bubble wall velocity, and $\Gamma$ the adiabatic index.
The spectral shape is similar to that mentioned in the previous section,
\begin{equation}
S_\text{sw}(f, f_0) =  \left(\frac{f}{f_0}\right)^3  \left(\frac{7}{4 + 3 (f/f_0)^2 } \right)^{7/2}
\end{equation}
where the peak frequency $f_0$ is
\begin{equation}
  \label{eq:peak_freq}
  f_0 = 8.9 \, \mu\mathrm{Hz} \,
  \frac{1}{v_\mathrm{w}} \left( \frac{\beta}{H_*} \right) \left(
  \frac{z_\mathrm{p}}{10} \right) \left( \frac{T_*}{100 \,
    \mathrm{GeV}}\right) \left( \frac{g_*}{100} \right)^\frac{1}{6}.
\end{equation}
Note that, in line with the general rules of thumb we developed above, the power
goes as $f^3$ at long wavelengths; and \eqref{eq:peak_freq} is approximately
Eq.~\ref{eq:generic_peak_freq} with the scale $B$ set by $R_*$, the bubble
radius at collision and assuming $v_\mathrm{w}\approx 1$, and the factor
$z_\mathrm{p} = 10$. More subtle predictions are being developed, based on the
assumption that the gravitational waves come entirely from overlapping sound
shells -- the `sound shell model'~\cite{Hindmarsh:2016lnk,Hindmarsh:2019phv}.
For example Refs.~\cite{Sharma:2023mao,RoperPol:2023dzg} show that below the
peak there is a wide `plateau' with a relatively flat $f^1$ power spectrum,
before rising sharply to a peak with an $f^9$ power law. These results have been
validated by simulations.

In any case the above expression, \eqref{eq:gwansatz}, and the sound shell
modelling results, are predicated on the assumption that $\tau_{\text{nl}}$ in
\eqref{eq:nltimescale} is longer than the Hubble time $1/H_*$, so that the
source is diluted by the expansion of the universe before nonlinearities start
to matter. If that is not the case, then the ans\"atze above may no longer
apply. How exactly they should be changed to produce accurate predictions in
general remains an open research problem, but good progress is being made in
understanding how the sound waves give rise to shocks and turbulence.

\subsection{Vacuum and runaway transitions}
\label{sec:vacuumrunaway}

Some scenarios, where the bubble wall moves very quickly, will not stir up
enough kinetic energy in the plasma to create a lasting source of gravitational
waves through sound waves. Some of these are so-called `runaway scenarios',
where there is friction between the bubble wall and the plasma, but it is
insufficient to slow the wall down. Others arise when the field that forms the
bubbles does not feel any plasma at all. In either case, the bubble walls
accelerate to ultrarelativistic speeds.

Most of the potential energy released by the transition is converted into
accelerating the bubble walls. The walls Lorentz contract at a rate which turns
out to be proportional to the ratio of the bubble radius to its value at
nucleation, $R_0$
\begin{equation}
  \gamma(R) = \frac{R}{R_0}.
\end{equation}
This, in turn, means that scalar field gradient energy is stored in the walls.
When the walls collide, a great deal of energy is released. Furthermore, with
limited damping, the field which underwent the transition can continue to
oscillate even after the bubbles have collided.

Because the walls get so severely Lorentz contracted, there is an even larger dynamic
range required between the wall width and the bubble radius in order to get
useful results from simulations, although there has been recent progress in
matching the power spectrum to the underlying
theory~\cite{Cutting:2018tjt,Cutting:2020nla}.

Many attempts have been made to analyse these vacuum and runaway scenarios
analytically. There is at the moment limited agreement between simulation and
analytical models, however one of the most promising is the so-called `bulk flow
model'~\cite{Konstandin:2017sat,Jinno:2017fby}.

\section{Detection}
\label{sec:Detection}

Supposing, then, that one or more phase transitions did happen in the early
universe, and they they did produce a cosmological stochastic background of
gravitational waves. How would we go about detecting them? Or, conversely, if we
understand the mechanisms of gravitational wave production sufficiently well,
can we use gravitational waves to rule out certain theories or scenarios?

As we will show in this section, we will most likely need to go into space to
detect a cosmological stochastic gravitational wave background, and missions
like LISA will be our best chance. However, there are also indirect constraints
on gravitational waves set by the cosmic microwave background and astrometric
observations, as well as evidence for a gravitational wave background at
nanohertz frequencies seen by pulsar timing arrays.

\subsection{Direct detection of gravitational waves}

In 2015 the Laser Interferometer Gravitational-Wave Observatory detected a
gravitational wave signal for the first time. Arising from the merger of two
black holes at about 410 megaparsecs, this and subsequent discoveries have
rekindled interest in direct detection of gravitational waves as a probe of
cosmology and astrophysics.

The two detectors of LIGO, and other Earth-based missions including VIRGO and
KAGRA, all function as very large, efficient Michelson interferometers.

\subsubsection{Detecting a gravitational wave with a Michelson interferometer}
\label{sec:Michelson}

\begin{figure}
    \centering
    \includegraphics[width=0.6\textwidth]{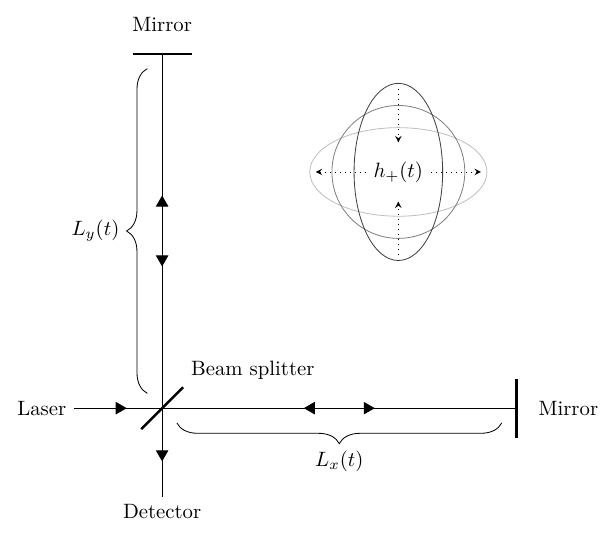}
    \caption{ Schematic depiction of a Michelson interferometer for the
    detection of gravitational waves. The lengths $L_x$ and $L_y$ of the two
    interferometer arms can be thought of as time-varying, although a proper
    analysis requires computing the time taken for photons to travel from the
    beam splitter to each mirror and back again. The interferometers used by the
    LIGO/Virgo/KAGRA collaboration follow this general principle. } 
    \label{fig:michelson}
\end{figure}
    
In a Michelson interferometer, a beam splitter sends coherent light down two
perpendicular arms, at the end of which they are reflected by a mirror. When the
light is recombined, the two split beams interfere (see
Fig.~\ref{fig:michelson}). The interference can be constructive, or destructive,
depending on the relative path lengths.

Following Ref.~\cite{Maggiore:2007ulw}, let us suppose we build such an
interferometer, and a gravitational wave passes through it perpendicular to the
plane in which it lies. For simplicity, let us assume that the wave is fully
polarised in the same directions as the arms:
\begin{equation}
  h_\times(t) = 0; \qquad h_+(t) = h_0 \cos(\omega_\text{gw} t)
\end{equation}
where $\omega_\text{gw}$ is the angular frequency of the gravitational wave and
$h_0$ is the strain amplitude. We will think of this as just a local stretching
of spacetime and neglect all other gravitational effects on the detector.

We can then write down the metric, or infinitesimal interval, $ds^2$, noting
that photons obey $ds^2=0$,
\begin{equation}
  ds^2 = c^2 \mathrm{d}t^2 + \left[ 1+h_+(t) \right]\mathrm{d}x^2 +
  \left[1 - h_-(t) \right]\mathrm{d}y^2 + \mathrm{d}z^2.
\end{equation}
We can use this to calculate the round-trip time for a photon down each arm, and
thus the phase shift. Normally the Michelson interferometer is set up to
destructively interfere when no gravitational waves are present, and so we are
looking to get the largest signal out when our gravitational wave passes
through. For a given angular frequency $\omega_\text{gw}$, this comes from
requiring
\begin{equation}
  \frac{\omega_\text{gw} L}{c} = \frac{\pi}{2} \Rightarrow L = \frac{\lambda_\text{gw}}{4}.
\end{equation}
In other words, for gravitational waves around $100\,\mathrm{Hz}$ (such as those
produced by stellar mass black hole binaries), we would need a
$750\,\mathrm{km}$ detector. Luckily, advances in optics have meant that, by
replacing the arms with Fabry-P\'erot cavities, we can get this down to a few
kilometres. On the other hand, if we want to detect gravitational waves with
frequencies around $1\, \mathrm{mHz}$, then something entirely different will be
required.

\subsubsection{Time-delay interferometry in space}

Space-based interferometer experiments such as the Laser Interferometry Space
Antenna (LISA) and proposals such as DECIGO and TianQin are in many ways similar
to their ground-based siblings. As on earth, these experiments take advantage of
phase difference between laser beams to measure perturbations in the
gravitational field. But space-based experiments can have much longer arms, and
are therefore sensitive to much lower frequencies (and larger wavelengths). In
the case of LISA, the arm-lengths will be approximately 2.5 million kilometres,
such that the gravitational wave frequencies probed can be estimated using 
\begin{equation}
    f_\text{gw} = c/ L = \frac{3.0 \times 10^5 \,{\rm km/s}}{ 2.5 \times 10^6 \,{\rm km}} = 0.12 \, {\rm Hz}.
\end{equation}
In practice this is an overestimate; LISA is most sensitive around $3 \, {\rm
mHz}$ due to the anticipated noise sources and detector response.

Another difference is that it is impossible to maintain an equal distance
between three spacecraft orbiting the sun, which means that the Michelson
interferometer with one laser and two mirrors cannot be used. Furthermore, the
constellation of the three spacecraft are rotating about the centroid of the
triangle shape. The experiment could therefore naively be much more sensitive to
laser noise with amplitudes far exceeding expected gravitational wave signals --
by as many as ten orders of magnitude. The solution to this problem is
time-delay interferometry~\cite{1999ApJ...527..814A}: if each of the spacecraft
of a space-based interferometer can both receive and transmit laser signals, the
laser noise from one transmitter can be measured by the others. 

\begin{figure}
  \centering
  \includegraphics[width=0.95\textwidth]{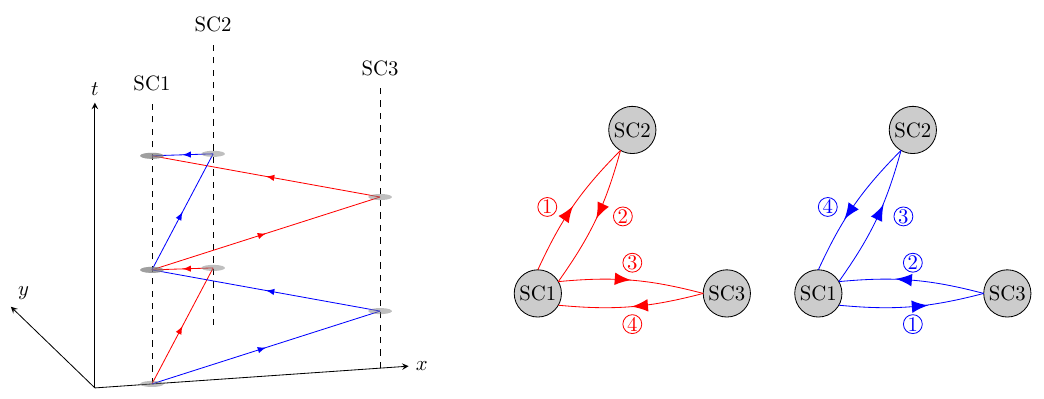}
  \caption{Schematic illustration of time delay interferometry, after Figure 2
  in Ref.~\cite{Hartwig:2022yqw}. Three spacecraft (`SC1', `SC2' and `SC3') are
  depicted (at left) in some choice of reference frame, while (at right) two
  plan views of the three spacecraft show the order in which laser signals are
  combined to carry out measurements. Laser signals are sent from one spacecraft
  and then received by another, forming a one-way measurement. The simplest
  combination that can be used to detect gravitational wave signals (a
  first-generation Michelson combination) is shown here: starting at SC1, one
  set of signals (red) involves measuring the delay to SC2 first and then SC3,
  while the other (blue) measures the delay to SC3 first and then SC1. These
  combinations can be thought of virtual laser beams, which interfere when they
  return to SC1. The combination of measured delays is not transmitted between
  spacecraft, it is instead constructed from the delay data afterwards.
  \label{fig:tdi}
  } 
\end{figure}

The laser noise of the receiving spacecraft enters the Doppler data immediately
at the time of reception, while the laser noise of the transmitting spacecraft
enters at a one-way delay time earlier. Because each spacecraft is transmitting
laser light at the two other spacecraft, there are six such one-way measurements
in the LISA constellation. Then, time-delayed combinations of these six signals
can be compared to cancel out the laser noise while retaining gravitational wave
signals (see Figure~\ref{fig:tdi}). In practice, these combinations can get
quite elaborate, but are sufficient to bring the laser noise down to acceptable
levels.

\subsubsection{Atom interferometry}
A different technique (see Ref.~\cite{Buchmueller:2023nll} for a recent
accessible summary) to study low frequencies takes advantage of the fact that
matter can behave like waves, especially at low temperatures. Particles such as
atoms have a de Broglie wavelength 
\begin{equation}
    \lambda_{\rm dB} = \frac{h}{mv}
\end{equation}
where $h$ is Planck's constant. From this it is seen that the smaller the
momenta of the particles (the lower their temperature), the longer the
wavelength. In atom interferometers, ultra-cold atoms are accelerated using
electromagnetic fields, split and brought back together to study the
interference pattern just like in laser interferometry. When gravitational waves
affect the travelling atoms, the interference pattern will be affected. 

An important difference is that atoms are affected by the gravitational pull of
the Earth much more than photons are. To account for this effect, atom
interferometers usually have vertical beams in which the atoms are brought into
free-fall. The frequency the experiments are sensitive to depends on the amount
of time free-fall can be realised; in the 100 m shaft of the MAGIS-100
experiment for example, the atoms can be in free-fall for $\sim 3 \; \rm s$ and
therefore the sensitivity is $f \sim 1/3 \; \rm Hz $.

\subsubsection{Proposals to detect high frequency gravitational waves}
Detecting high-frequency GWs is an active area of research in astrophysics and
gravitational physics, aiming to open new windows into the universe beyond what
has been possible with current low-frequency GW observatories like the ones
described above. High-frequency gravitational waves, typically considered to be
in the range of $10^4$ Hz to $10^9$ Hz, are challenging to detect with Michelson
interferometers as the thermal and quantum noise in this frequency window is too
large to make detections. 

Various proposals are being explored to achieve this, including electromagnetic
proposals primarily inspired by axion detectors (see e.g.
\cite{Domcke:2023qle}), resonant mass detectors that rely on the vibration of
solid objects (see e.g. \cite{Aguiar:2010kn} for a review), and opto-mechanical
devices that detect gravitational waves through changes in optical properties
\cite{Arvanitaki:2012cn}.\footnote{For an extensive review on all proposals to detect high-frequency gravitational waves and
further references, please refer to the living review \cite{Aggarwal:2020olq}.}
The direct detection of high frequency gravitational waves faces immense
challenges, owing to the small wavelength and large noise. Accordingly, the
sensitivity of these experimental proposals is not typically strong enough to
detect early Universe gravitational wave backgrounds from before the electroweak
epoch.

\subsection{Indirect detection of gravitational waves}
Besides the dedicated experiments measuring gravitational waves directly, their
presence can also be inferred from a number of indirect probes. Vice versa,
non-observation can be used to derive constraints.  

\subsubsection{Early universe probes}
The polarisation of the photons comprising the CMB can be expressed in a basis
of E-modes and B-modes. E-mode polarisation is in or perpendicular to the wave
vector, B-mode polarisation is at a $\pm 45 \degree$ angle. Density
perturbations are a scalar quantity and only generate E-mode polarisation, but
gravitational waves are tensors and generate both. B-modes have not been
observed by CMB experiments such as Planck and BICEP2
\cite{BICEP2:2015nss,Planck:2018jri}, which are sensitive to frequencies in the
range $10^{-20}-10^{-16}$ Hz. Cosmic inflation itself leads to a gravitational
wave spectrum which can be well approximated by a power-law $ \Omega_{\rm GW}
\propto (f/f_{\rm cmb})^{n_t}$, with a small negative spectral index ${n_t}$ in
the simplest models. Constraints are often reported in terms of the
tensor-to-scalar ratio $r=A_t/A_s$, which measures the amplitude of the
gravitational wave spectrum in terms of that of the scalar perturbations. The
current constraint is $r < 0.12$, which corresponds to $\Omega_{\rm GW} \lesssim
10^{-14} $ at the pivot frequency $ f_{\rm cmb}\sim 10^{-17}$ Hz. The next
generation of CMB experiments, CMB stage-4, will improve upon this. 

Any new form of energy in the universe contributes to its expansion rate, as
captured by the Hubble constant. Non-relativistic or matter-like contributions
are different from relativistic or radiation-like contributions, and both the
CMB and measurements of the light elements produced in BBN can be used to
determine quite how much relativistic energy there was during the BBN epoch. As
gravitational wave energy is relativistic, this gives an upper bound on the
total amount of gravitational waves produced in the early universe. This amount
is often captured in terms of the effective number of neutrino species: i.e. how
many neutrino-like particles there could be if they accounted for the radiation:
\begin{align}\begin{split}
  \frac{\Omega_{\rm GW,0} h^2}{\Omega_{\gamma,0} h^2} &= \frac{1}{\Omega_{\gamma,0} h^2}
\int {\rm d} \log k \, \frac{1}{\rho_{\rm tot,0}} \dd{\rho_{\rm GW,0}}{\log k} = \frac{7}{8} \left(\frac{4}{11} \right)^{4/3} \Delta N_{\rm eff} \end{split}, 
\end{align}
where the $0$ denotes a present day quantity, and the present photon density is
given by $\Omega_{\gamma} h^2 = 2.47 \times 10^{-5}$. At the moment, the limit
is given by $ \Delta N_{\rm eff} = N_{\rm eff} - N_{\rm eff,SM} \leq \mathcal{O}
(10^{-1})$ \cite{Planck:2018vyg}, meaning that the integrated gravitational wave
energy today is already constrained to be smaller than $\mathcal{O} (10^{-7}) $.
Next-generation experiments will push down further, to about $ \Delta N_{\rm
eff} \leq \mathcal{O}(10^{-2})$, bringing the constraint down even more.

\subsubsection{Pulsar timing arrays}
Observations of the electromagnetic spectrum coming from astrophysical objects
can also encode information about gravitational waves. Millisecond pulsars are
an example. These rapidly spinning neutron stars have a period as stable as an
atomic clock, and therefore serve as a sensitive probe of their environment. The
presence of gravitational waves affects the pulses of EM radiation as they
travel towards the Earth; careful monitoring of the pulse arrival times can
therefore be used to detect or constrain gravitational waves. Pulsar timing
arrays (PTAs) do just that; these dedicated experiments typically monitor a
large number of pulsars weekly ($\sim 10^{-6}$ Hz) and run for a number of years
($\sim 10^{-8}$ Hz). This cadence is chosen primarily for the discovery of
galactic binary mergers, in which galactic nuclei -- supermassive black holes --
play the most important role. But other signals, such as that generated by a
first order phase transition at a relatively low scale could also fall within
this frequency window. Currently, PTAs can probe microhertz gravitational waves
with amplitude of about $\Omega_\text{GW} h^2 \sim 10^{-10}$ at 3~$\mu\mathrm{Hz}$~\cite{NANOGrav:2020bcs}.

In 2020, the NANOGrav collaboration reported a signal that could be consistent
with a stochastic gravitational wave background in the nanohertz range. This
signal was observed in the timing residuals of an array of millisecond pulsars
over a 12.5-year period. The detection was subsequently confirmed by other
experiments: the European Pulsar Timing Array (EPTA), the Parkes Pulsar Timing
Array (PPTA) in Australia, and the International Pulsar Timing Array (IPTA),
which is a consortium of all these major groups. For a PTA signal to be
confirmed as a gravitational wave background, and specifically as a signature of
early universe phase transitions, subsequent data must reveal the Hellings-Downs
correlation: this would show the characteristic quadrupolar spatial correlation.
This indicates that the observed patterns are indeed due to the space-time
ripples caused by gravitational waves, as opposed to any other confounding
signals or noise. Evidence for this correlation was reported simultaneously in
June 2023 by NANOGrav~\cite{NANOGrav:2023gor}, the Chinese Pulsar Timing
Array~\cite{Xu:2023wog}, EPTA and the Indian Pulsar Timing
Array~\cite{EPTA:2023fyk} and PPTA~\cite{Reardon:2023gzh}, with all of these
results apparently in agreement with one
another~\cite{InternationalPulsarTimingArray:2023mzf}.

Since the PTA signal was reported, several studies have analysed the data. Some
of these have found that early Universe signals such as first order phase
transitions provide a better fit to the signal than supermassive black hole
mergers (e.g.~\cite{Ellis:2023oxs,Wu:2023hsa}), based on basic modelling of both
of these signals, though others showed inconclusive evidence or a preference for
another source~\cite{Figueroa:2023zhu,Bian:2023dnv}. Such a phase transition
would have taken place at a temperature of the early Universe bath below the QCD
scale (see Table~\ref{tab:GWEvents} and Fig.~\ref{fig:Cosmictimeline}), but
above the scale of BBN ($T \sim 2\, \mathrm{MeV} $), or otherwise light element
abundances would be disrupted. A hidden sector, only gravitationally connected
to the Standard Model, could have sourced the nanohertz GWs. Such a phase
transition could also be supercooled, meaning that the universe remained in a
metastable state longer than expected. This delay allows the bubbles of the new
phase to grow larger and contain more energy, leading to stronger gravitational
wave signals. The increased bubble size can also mean that the peak frequency of
the emitted GWs is redshifted into the PTA-sensitive range.

A more definite answer on the source of the PTA signal may come from LISA, the
space interferometer discussed in the previous section. Because LISA probes
gravitational waves in a complementary frequency range, it might reveal more of
the spectral shape of the signal, which could then be used to distinguish between
the astrophysical and cosmological interpretations.

\subsubsection{Astrophysical probes}

\begin{figure}
    \centering
    \includegraphics[width=0.5\textwidth]{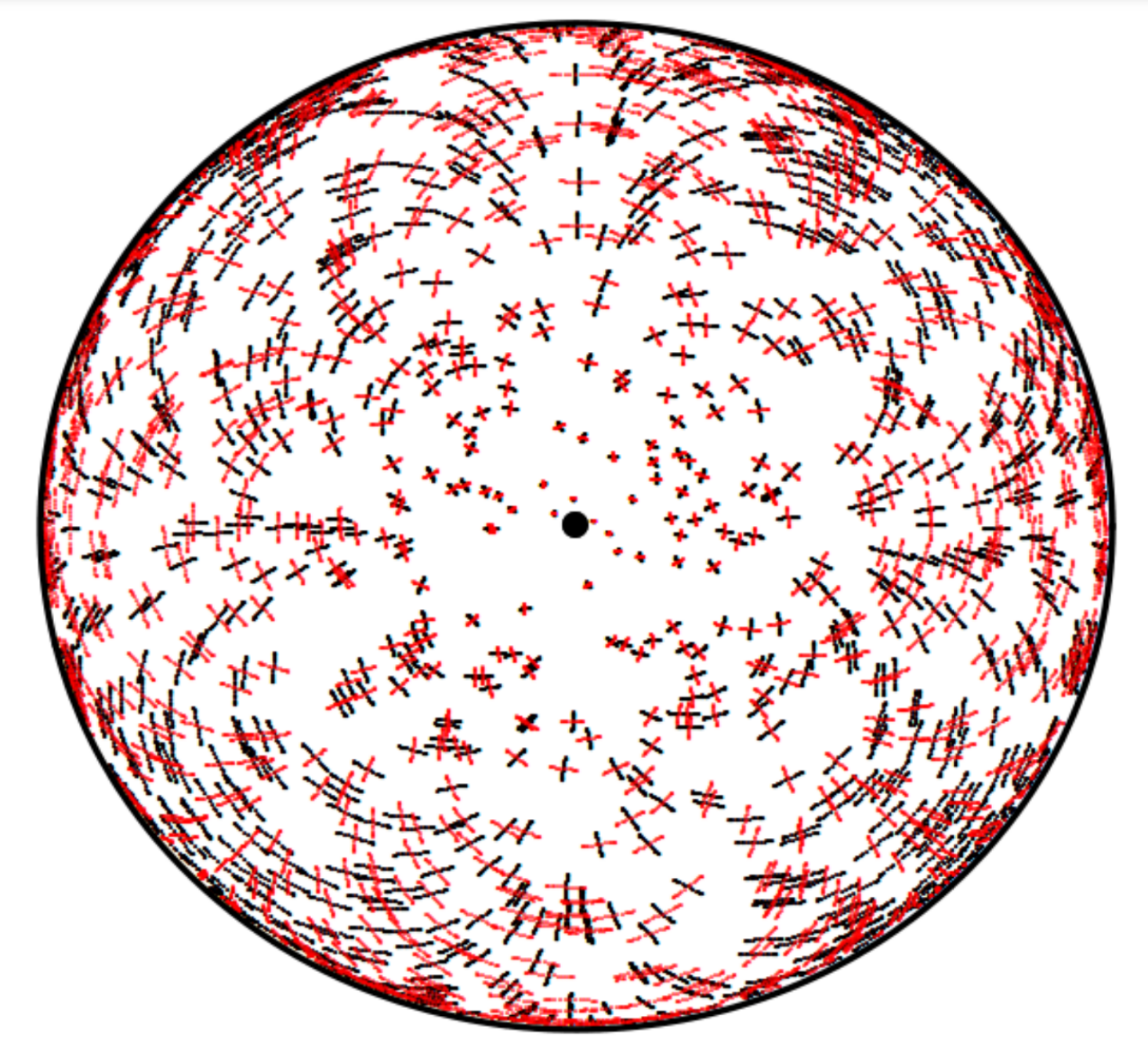}
    \caption{ Demonstration of astrometry as a probe of gravitational waves. The
    image shows an orthographic projection of the Northern hemisphere with 103
    stars, with a GW from the North pole (black dot) which causes the apparent
    position of stars to oscillate. The black (red) lines show movement tracks
    for a linearly plus (cross) polarisation. The GW amplitude has been
    exaggerated for clarity. Figure from \cite{Moore:2017ity}, used with
    permission.}
    \label{fig:astrometry}
\end{figure}

Precise astrometric observations of the motion of galactic stars can also be
used to constrain gravitational waves. Galaxy surveys such as GAIA
\cite{Gaia:2016zol} make precise observations of $ \sim10^9$ stars in our
galaxy. Gravitational wave signals would interfere with the light of those stars
as it travels towards the Earth, which would register as very small wiggles in
their apparent position -- see Fig.~\ref{fig:astrometry}. Because statistical
data analysis on the positions of $10^9$ objects is computationally unfeasible,
averages of sample averages can be taken without loss of sensitivity.
Astrometric observations are sensitive to gravitational waves in a similar
frequency range as PTAs, with similar sensitivity. This corresponds to distances
of about a parsec in apparent shift: 
$
    f_{\rm GW} \sim c/1 \; {\rm pc} \sim 10^{-9} \; {\rm Hz}
$.

\section{Conclusions and Outlook}
\label{sec:Conclusions}
The prospect of observing gravitational waves with millihertz frequencies serves
as a strong motivation to study out-of-equilibrium dynamics in the very early
universe. First order phase transitions -- events which have been studied in
connection with the matter/antimatter asymmetry, dark matter, and confinement --
give rise to stochastic gravitational wave backgrounds. In this review we have
discussed the particle physics behind a first order phase transition, how the
subsequent dynamics gives rise to gravitational waves, and how the various
experiments looking for gravitational waves work. 

The particle physics models studied for their phase transitions can typically
also be studied in other ways. For example, matter/antimatter asymmetry models
which use the conditions of the electroweak phase transition usually also
predict new particles at energy scales within reach of the next generation of
particle colliders. Therefore, complimentary studies can inform gravitational
wave searches, and vice versa: if a hint of new physics is found in the
gravitational wave data, follow up experiments will have to be done to verify
more details of its nature. 

For scientists interested in this research program, there is a lot of
interesting work ahead. In the next two decades, many new experiments will start
to produce data. To accurately find the implications of this data for particle
physics, our techniques for calculating and modelling the phenomenology need to
be improved. On the microphysics side, calculating the false vacuum decay rate
and the subsequent expansion of critical bubbles suffers from theoretical
uncertainty, related to finite temperature effects. On the computational side,
studying the onset of turbulence in a relativistic plasma is a very big but
important challenge. 

\paragraph{Acknowledgments:} D.C. is supported by the STFC under Grant No.
ST/T001011/1. D.J.W. was supported by Research Council of Finland grant no.
324882.

\bibliographystyle{apsrev4-1}
\bibliography{references}

\end{document}